\documentclass[review]{elsarticle}
\usepackage{geometry}
\usepackage{pdflscape}
\usepackage{tabularx}
\usepackage{makecell}
\usepackage{multicol}
\usepackage[utf8]{inputenc}
\usepackage[T1]{fontenc}
\usepackage[scaled=0.85]{beramono}
\usepackage{float}
\usepackage{graphicx}
\usepackage{subcaption}
\usepackage{booktabs, multicol, multirow}
\usepackage{makecell}
\usepackage{comment}
\usepackage{multirow}
\usepackage{graphicx}
\usepackage{lscape}
\usepackage{adjustbox}
\usepackage{multirow}
\usepackage{longtable}

\usepackage{multirow}
\usepackage{graphicx}
\usepackage[english]{babel}
\usepackage[scaled=0.85]{beramono}
\usepackage{amsmath}
\usepackage{stackengine}
\usepackage[colorlinks=true, allcolors=blue]{hyperref}
\usepackage{listings}
\usepackage{xcolor}
\usepackage[colorinlistoftodos]{todonotes}

\definecolor{codegreen}{rgb}{0,0.6,0}
\definecolor{codegray}{rgb}{0.5,0.5,0.5}
\definecolor{codepurple}{rgb}{0.58,0,0.82}
\definecolor{backcolour}{rgb}{0.95,0.95,0.92}

\lstdefinestyle{mystyle}{
   backgroundcolor=\color{backcolour},   
    commentstyle=\color{codegreen},
    keywordstyle=\color{magenta},
    numberstyle=\tiny\color{codegray},
    stringstyle=\color{codepurple},
    basicstyle=\ttfamily\footnotesize,
    breakatwhitespace=false,         
    breaklines=true,                 
    captionpos=b,                    
    keepspaces=true,                 
    numbers=left,                    
    numbersep=5pt,                  
    showspaces=false,                
    showstringspaces=false,
    showtabs=false,                  
    tabsize=2
}

\usepackage{lineno,hyperref}

\begin{document}

\begin{frontmatter}

\title{\LARGE \bf
Bangladesh Agricultural Knowledge Graph: Enabling Semantic Integration and Data-driven Analysis--Full Version}

\author[mymainaddress]{Rudra Pratap Deb Nath\corref{mycorrespondingauthor}}
\cortext[mycorrespondingauthor]{Corresponding author}
\ead{rudra@cu.ac.bd}

\author[mymainaddress]{Tithi Rani Das}
\ead{tithi.rani.das.cucse@gmail.com}   
\author[mymainaddress]{Tonmoy Chandro Das}
\ead{tonmoycsecu@gmail.com} 

\author[mymainaddress]{S. M. Shafkat Raihan}
\ead{shafkatraihan9001@gmail.com}    

\address[mymainaddress]{Department of Computer Science and Engineering, University of Chittagong, Chattogram-4331, Bangladesh}

\begin{abstract}
In Bangladesh, agriculture is a crucial driver for addressing Sustainable Development Goal 1 (No Poverty) and 2 (Zero Hunger), playing a fundamental role in the economy and people's livelihoods. To  enhance the sustainability and resilience of the agriculture industry through data-driven insights, the Bangladesh Bureau of Statistics and other organizations consistently collect and publish agricultural data on the Web. Nevertheless, the current datasets encounter various challenges: 1) they are presented in an unsustainable, static, read-only, and aggregated format, 2) they do not conform to the Findability, Accessibility, Interoperability, and Reusability (FAIR) principles, and 3) they do not facilitate interactive analysis and integration with other data sources. In this paper, we present a thorough solution, delineating a systematic procedure for developing BDAKG: a knowledge graph that semantically and analytically integrates agriculture data in Bangladesh. BDAKG incorporates multidimensional semantics, is linked with external knowledge graphs, is compatible with OLAP, and adheres to the FAIR principles. Our experimental evaluation centers on evaluating the integration process and assessing the quality of the resultant knowledge graph in terms of completeness, timeliness, FAIRness, OLAP compatibility and data-driven  analysis.  Our federated data analysis recommend a strategic approach focused on decreasing CO$_2$ emissions, fostering economic growth, and promoting sustainable forestry.
\end{abstract}


\end{frontmatter}


\section{Introduction}\label{sec:intro}

The heyday of civilization was greatly aided by agriculture. Its significance in the economy of any nation, whether developing, undeveloped, or developed, is incalculable and ranges from industrial production to national wealth.  Nowadays, agriculture has substantially evolved with the advent of data-driven decision-making. This approach involves the collection, integration, publication, analysis, and interpretation of various types of data to inform and optimize agricultural practices. By harnessing data, farmers and stakeholders in the agricultural sector can make more informed and precise decisions, leading to improved productivity, resource management, and sustainability.

In the socio-economic context of Bangladesh, particularly as a developing nation, agriculture 1) serves as the primary pillar of its economic foundation and livelihood,  and 2) plays a crucial role in addressing two  of the United Nations Sustainable Development Goals (SDGs): SDG 1 (No Poverty) and SDG 2 (Zero Hunger). To shape a more sustainable and resilient future for this industry through advanced technology and data-driven insights, Bangladesh government and related organizations emphasize on the synergy between agriculture and its digital twin~\citep{pylianidis2021introducing}. 
Following this trend, the Bangladesh Bureau of Statistics (BBS) \cite{Bangladesh:online} and other organizations regularly manage and publish agricultural  data   comprising of  crops, fisheries, livestock, and forestry on the Web.

In order to fully grasp the insights within these published data, it is essential that the presentation of these data should be not only  in an analytical format but also in a findable, accessible, compatible, and sustainable format, enabling the data to be repurposed and interconnected across various sources~\cite{deb2022high}. However, a prevailing practice  is that most of the agriculture data sources presents data in an unsustainable (become unavailable before outdated), static, and read-only (pdf) format, rather than for interactive analysis or integration with other data sources. The limitations of those sources are as follows: 1) They are not following the Findability, Accessibility,
Interoperability, and Reusability (FAIR) principles~\cite{jacobsen2020fair}. Hence, they are difficult for researchers and other interested parties to discover, integrate with other data and (re)use in new context or for new purposes. 2) Different
sources describe the same data in different ways, introducing semantic heterogeneity problems.
3) Most of the data are published at an aggregate level, which does not present complete understanding of complex issues and phenomena.  4) Data are not globally interlinked, therefore, deriving insights by analyzing or comparing with different datasets is not supported.


The constraints outlined regarding the data curtail its potential worth, utility, and influence. The true potential of the data is realized when they are 1) semantically defined, 2) efficiently integrated, and 3) analytically explored. Here, semantically defined means data are clearly and explicitly specified. Efficient integration means data from multiple sources are semantically integrated in a single source of truth. Finally, analytically exploration enables descriptive---Online Analytical Processing(OLAP)-like--- and exploratory analysis on the integrated dataset.


In this context, the Semantic Web (SW)~\cite{antoniou2004semantic} emerges as a solution. The SW  annotates web-published content with machine-understandable, semantic information so it may be efficiently retrieved and processed by both people and machines in a wide range of tasks~\cite{nebot2012building}. The SW encourages organizations to arrange and publish data following a set of design principles, known as Linked Data (LD) principles  \cite{bizer2011semantic}.  By adhering to these principles and representing data in Resource Description Framework (RDF)~\cite{decker2000semantic}, the LD approach allows for seamless integration of data from different sources. It models the data as a knowledge graph~\cite{chen2020review}, enabling users and applications to discover, navigate, and query a vast network of interconnected information. In this way, it fulfills the FAIR principles and promotes the creation of a more coherent and comprehensive information space on the Web.  

 To unlock the inherent possibilities present in the open agricultural data of Bangladesh, this study follows three steps: 1) Modeling Bangladesh agriculture open data semantically and analytically, 2) Integrating data semantically from diverse sources, conforming to the model, and publishing data as LD and FAIR data, and 3) Enabling exploratory analytics over the integrated data. Firstly, we discover and collect the data related to crops, forestry, fisheries, etc. from different open sources enlisted in Table~\ref{tab:soucedataset}. Then, we follow the demand-driven approach~\cite{lenzerini2002data} to model the data. For capturing data at schema level, we  use an ontology as it provides a formal and expressive means of representing the structure and semantics of data within a domain~\cite{gomez2002ontology}. Since the dataset should also be presented in an analytical format to enable OLAP-style analysis, we use QB and QB4OLAP vocabularies to annotate the ontology with Multidimensional (MD) semantics. Once data are modeled, then the ontology is populated from different sources creating a semantic Extract-Transform-Load (ELT) flow. We name the integrated datasets (Ontology along with its instances) as BDAKG: Bangladesh Agricultural Knowledge Graph. For making BDAKG more exploreable, it is semantically linked with other external knowledge graphs. Moreover, because of annotating with MD semantics, BDAKG enables OLAP queries, where data cubes are explored through user-friendly interfaces. 
 
 We summarize the novel contributions of this paper as follows.

 \begin{itemize}
     \item[-] We create  BDAKG: a semantic repository or knowledge graph for Bangladesh agriculture open data, by semantically integrating different sources and annotating them with MD semantics. 
     \item[-] We outline the comprehensive and step-by-step data integration process. 
     \item[-] We link BDAKG internally and externally with other datasets available in Linked Open Data (LOD) cloud to enable exploratory analytics.
     \item[-] We assess both the integration process and BDAKG concerning the data quality standards, encompassing data FAIRness, OLAP-compatibility, correctness, and federated data analysis capability. Our data-driven analysis suggests a course of action aimed at reducing CO$_2$ emissions, promoting economic growth, and fostering sustainable forestry. 
 \end{itemize}

The remainder of the paper is organized as follows. We discuss the notations and terminologies used throughout the paper in Section~\ref{sec:pd}. Section~\ref{sec:uc}
details the source datasets and the ontology of the BDAKG. Section~\ref{sec:md} describes the process of creating BDAKG. 
We describe BDAKG and evaluate it in terms of performance, quality, and federated data analysis in Section~\ref{sec:exp}. We position our research contribution within the current state-of-the-art, as detailed in Section~\ref{sec:rw}.. Finally, we conclude and give pointers to future work in Section~\ref{sec:con}.

\section{Materials and Methods} In this section, we first introduce the fundamental concepts and terminologies used in this paper. Following that, we outline the source datasets and the conceptual model of the data intended for integration. Subsequently, we elaborate on the methodology employed for generating BDAKG.

\subsection{Preliminaries } \label{sec:pd}

\begin{figure*}[h!]
\centering
\includegraphics[scale=.7]{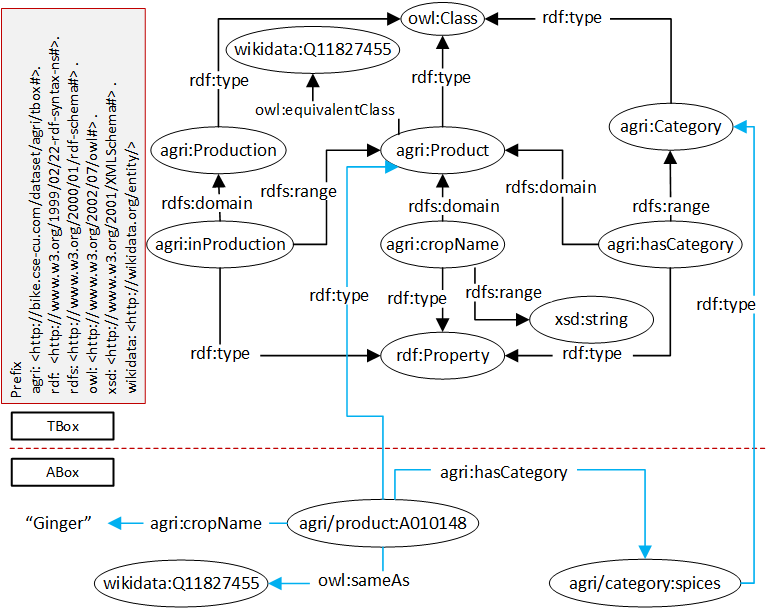}
    \caption{ A knowledge graph presenting a portion of TBox and ABox.}
    \label{fig:kg}
\end{figure*}

 Here, we define the key concepts and terminologies used throughout the paper.

\subsubsection{Knowledge graph}
 A knowledge graph (KG) is a directed graph designed to accumulate and communicate information about the real world. In this graph, nodes symbolize entities of interest, while edges signify the connections or relationships between these entities. A KG consists of two components: the Terminological Box (TBox) and the Assertion Box (ABox). The TBox establishes domain-specific terminology, while the ABox contains assertions representing individual instances. It is important for the ABox assertions to follows the semantics encoded in the TBox~\cite{deb2022high}. In this paper,  we express KG components using a collection of RDF triples. An RDF triple is formally defined as a 3-tuple $(s,o,p)$, where $ s\in (I\cup B)\;, p\in I,$ and $o \in I \cup B \cup L$, and a KG is defined as \( KG \subseteq (I\cup B)\times I \times(I\cup B \cup L)\). Here, \(I\), \(B\), and \(L\) stand for the sets of IRIs, blank nodes, and literals, respectively. An IRI is a unique identifier for globally identifying a resource (Web-scope). Blank nodes function as locally-scoped identifiers for resources unknown to the external world. Literals are lexical values enclosed in inverted commas. 
 
 We use basic constructs from formal languages such as RDF-schema (RDFS) and the Web Ontology Language (OWL) in conjunction with the RDF data model to impose richer constraints on data. Figure~\ref{fig:kg} shows a portion of BDAKG presenting its components. The labeled vertices depict subjects and objects in RDF triples, directed labeled edges represent predicates, and literal-vertices are represented by plain text. A red-dotted line serves to distinguish the TBox and ABox.  In practice, a KG is represented as an RDF graph without differentiating between classes and instances. 
The figure annotates \texttt{agri:Product}, \texttt{agri:Production}, and \texttt{agri:Category} as classes, and the properties \texttt{agri:inProduction}, \texttt{agri:cropName}, and \texttt{agri:hasCategory} are connected with the classes using the \texttt{rdfs:domain} and \texttt{rdfs:range} constructs. A class offers a broad overview of the characteristics shared by related resource types. Meanwhile, a property establishes connections between instances of a class or links instances to literal values. The ABox presents that \texttt{agri/product:A010148} is a product and its name and category is spices.  KGs can also be linked with external KGs (in this case, Wikidata~\cite{vrandevcic2014wikidata}) through  \texttt{owl:equivalentClass}, \texttt{owl:equivalentProperty} (at the TBox level) and \texttt{owl:sameAs} (at the ABox level) properties. Thus, KGs enable exploration through linking. 

\subsubsection{Multidimensional modeling and the QB4OLAP vocabulary}\label{sec:mmqv}
Agricultural data encompasses factual information and numerical values. Consequently, it is advisable to present the data in a format compatible with OLAP. OLAP technology plays a crucial role in facilitating data analysis to support decision-making processes. For effective OLAP queries, it is recommended to represent a KG in a Multidimensional (MD) model. The choice of the MD model is motivated by its simplicity and the straightforward, intuitive approach it provides for conducting analytical queries. 

In the MD model, data are viewed in an n-dimensional space, generally known as a data cube, composed of facts (the cells of the cube) and dimensions (the axes of the cube).  Hence, it enables users to examine data across various dimensions of relevance. For instance, a user can analyze production of products according to geographical regions and time (dimensions). Facts are the interesting things or processes to be analyzed (e.g., production of products, sales of product) and the attributes of the fact are called measures (e.g., production, area, amount of sales, quantity),
usually represented as numeric values. Dimensions are organized into hierarchies (composed of a number of levels) to explore and (dis)aggregate fact measures (e.g., numerical data) at various levels of detail~\cite{varga2016qb2olap}. For example, the \texttt{agri:GeographyDim} dimension’s hierarchy \texttt{agri:GeoHierarchy} $(District\rightarrow Division \rightarrow All)$ enables to explore and (dis)aggregate the agricultural production information of Bangladesh at various administrative levels of detail (see Figure~\ref{fig:schemaDiag1}).

To annotate BDAKG with MD semantics, we use the Data Cube for OLAP (QB4OLAP) vocabulary~\cite{varga2016qb2olap} which extends the Data Cube (QB) vocabulary~\cite{etcheverry2012qb4olap} to overcome QB's limitations in presenting the MD constructs. Table~\ref{tab:qb4olap} summarizes the different QB4OLAP constructs used to define corresponding MD constructs.

\begin{table}[h!]
\centering
\caption{QB4OLAP constructs used to define MD constructs.}
\label{tab:qb4olap}
\resizebox{\textwidth}{!}{%
\begin{tabular}{|l|l|l|l|}
\hline
\textbf{MD construct}                                            & \textbf{Mapped QB4OLAP construct}                  & \textbf{MD construct}                                        & \textbf{Mapped QB4OLAP construct}                                                                     \\ \hline
Dimension                                                         & \texttt{qb:DimensionProperty} & Dataset                                                       & \texttt{qb:DataSet}                                                            \\ \hline
Hierarchy                                                         & \texttt{qb4o:Hierarchy}     & Cube                                                          & \begin{tabular}[c]{@{}l@{}}\texttt{qb:DataStructureDefinition}\end{tabular} \\ \hline

\begin{tabular}[c]{@{}l@{}}Connections between\\ Dimensions and Hierarchies\end{tabular} & \texttt{qb4o:inDimension}, \texttt{qb4o:hasHierarchy} &  Fact                                                          & \texttt{qb:Observation}                                                   \\ \hline

Level                                                             & \texttt{qb4o:LevelProperty} &   Level Instance                                                     & \texttt{qb4o:LevelMember}                                                        \\ \hline

\begin{tabular}[c]{@{}l@{}}Connection among\\ Levels\end{tabular} & \texttt{qb4o:HierarchyStep} & Attribute                                                     & \texttt{qb4o:LevelAttribute}                                                   \\ \hline
Measures                                                          & \texttt{qb:MeasureProperty} & \begin{tabular}[c]{@{}l@{}}Aggregate \\ Function\end{tabular} & \texttt{qb4o:AggregateFunction}                                                \\ \hline
\end{tabular}%
}
\end{table}




\subsection{Dataset descriptions and the target TBox modeling}
\label{sec:uc}
In this section, we provide an overview of the source datasets and introduce the target TBox used to consolidate and understand the knowledge within these datasets.

\subsubsection{Description of Source Datasets }

We utilize an agricultural dataset obtained from the Bangladesh Bureau of Statistics (BBS)~\cite{Bangladesh:online}, specifically from the Yearbook of Agricultural Statistics of Bangladesh, which is published annually. The BBS has consistently served as a reliable source of information on crop production since 1974, covering six major crops and over a hundred minor crops. According to the Statistics Act of 2013~\cite{2022013137:online}, the BBS extensively provides data on the crop sub-sector, with a primary focus on crop data and limited information on the fisheries, forestry, and livestock sub-sectors. The yearbook presents comprehensive statistics that encompass the nation's agro-ecology, land types, soil classification, crop seasons, land use statistics, meteorological data (including rainfall, temperature, humidity, and major cyclones), agricultural inputs, livestock statistics, fisheries, forestry, crop prices, and export and import data on agricultural products and inputs. It also provides significant agricultural insights. We have organized the dataset into three main segments: crops, fisheries, and forestry. Table~\ref{tab:soucedataset} outlines a summary of these datasets.

\textbf{Crops Dataset:} For the crop dataset, we selectively consider data as measures that can be effectively analyzed based on three key dimensions: Time, Geography, and Product. From the yearbooks, we extract data for major crops (aus, aman, boro, jute, potato, and wheat) \cite{45yearsM5:online} spanning from the 1969–1970 to 2019–2020 periods. Additionally, we gather data for minor crops (barley, onion, ginger, carrot, and lalsak, etc.) \cite{Bangladesh:online} from 2007–2008 to 2019–2020, focusing on area (in acres) and production (in metric ton). We also compile information on each crop's harvest time and sowing time. Drawing insights from \cite{leff2004geographic, Banglade50:online}, we categorize all crops into 15 distinct categories such as cereals, pulses, seeds, etc., and provide comprehensive descriptions along with a unique key for easy identification of each crop category. The scientific names of the crops are sourced from various reliable websites \cite{zimdahl1989weeds}. 

\textbf{Fisheries Dataset:}
For the fisheries dataset \cite{Bangladesh:online}, we only extract production data with respect to fisheries habitat (river, estuary, beel, and pond, etc.) between 2010–2011 and 2019–2020.

\textbf{Forestry Dataset:}
We collect forestry data from the BBS \cite{Bangladesh:online} between 2010–2011 and 2017–2018.

\textbf{Other Datasets:}
For country, district and division levels, we collect all data, including unique codes, from  Division-of-the-sectors\cite{Division74:online}, O-Subentity-Codes-for-Bangladesh \cite{Administ1:online}.

\begin{table}[h!]
\centering
\caption{Overview of the data sources}
\label{tab:soucedataset}
\resizebox{\textwidth}{!}{%
\begin{tabular}{|l|l|l|l|}
\hline
\textbf{Dataset} & \textbf{Original format} & \textbf{Number of  instances} & \textbf{Data sources}                                                                                                                                                                                                                                                                \\ \hline
Crops            & Pdf, Text, Tabular       & 54712                         & \begin{tabular}[c]{@{}l@{}}BBS\cite{Bangladesh:online}, Classification-of-crops\cite{leff2004geographic},\\  Crops-code \cite{Banglade50:online}\end{tabular} \\ \hline
Fisheries        & Pdf, Text, Tabular       & 6743                          & BBS\cite{Bangladesh:online}, Fisheries \cite{shamsuzzaman2017fisheries}                                                                                                                                                                                  \\ \hline
Forestry         & Pdf, Text, Tabular       & 520                           & BBS\cite{Bangladesh:online}, Forestry\cite{helms2002forest}                                                                                                                                                                                    \\ \hline
Other Datasets   & Text, Tabular            & 73                            & \begin{tabular}[c]{@{}l@{}}Division-of-the-sectors\cite{Division74:online},\\ O-Subentity-Codes-for-Bangladesh \cite{Administ1:online}, Livestock \cite{thornton2010livestock}\end{tabular}                                                                                                \\ \hline
\end{tabular}%
}
\end{table}

\subsubsection{Target TBox definition}\label{sec:ttd}

\begin{figure*}[h!]
\centering
\includegraphics[scale=.42]{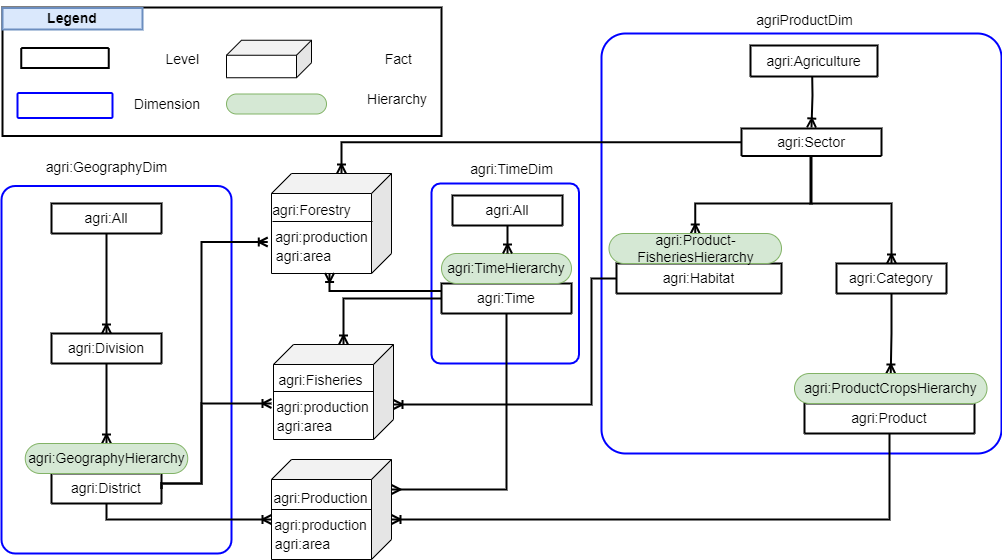}
    \caption{ The TBox of Bangladesh Agriculture Knowledge Graph (BDAKG). The figure does not display level attributes due to their substantial quantity.}
    \label{fig:schemaDiag1}
\end{figure*}

To integrate the agriculture datasets, we define a TBox of BDAKG. The TBox of BDAKG is depicted in Figure \ref{fig:schemaDiag1}. The data is organized in multidimensional (MD) structures resembling cubes, with dimensions spanning their axes. Rather than utilizing a single data cube, we employ three distinct cuboids—forestry, fisheries, and production—to formulate the fact constellation schema, where each level is normalized. We use two measures:  \texttt{agri:area} and \texttt{agri:production}, which are associated with each and every point in this MD space. In the figure, the cubic box and rectangles represent the fact table cubes and levels, respectively, while the blue rectangles represent the dimensions.

  The TBox has three dimensions: \texttt{agri:GeographyDim}, \texttt{agri:ProductDim}, and \texttt{agri:TimeDim}. Three dimensions are used to analyze the agriculture data from multiple perspectives, enabling OLAP operations. To describe the data being analyzed at various levels of abstraction, hierarchies \cite{vaisman2014data} are essential components of analytical applications. The \texttt{agri:GeographyDim} dimension’s hierarchy \texttt{agri:GeoHierarchy} $(\texttt{agri:District} \leftrightarrow \texttt{agri:Division} \rightarrow \texttt{agri:All})$ enables to explore and (dis)aggregate the agricultural production information of Bangladesh at various administrative levels of detail. The \texttt{agri:ProductDim} dimension has two hierarchies, namely \texttt{agri:ProductCropsHierarchy} and \texttt{agri:ProductFisheriesHierarchy}. Figure \ref{fig:schemaDiag1} describes the two aggregation paths, one for each type of crop and another for fisheries data, where both remain in the same hierarchy. Products must be aggregated differently according to the product's type; for crops, the aggregation path is $\texttt{agri:Product}\rightarrow \texttt{agri:Category} \rightarrow \texttt{agri:Sector} \rightarrow \texttt{agri:Agriculture}$; for fisheries, the path is $\texttt{agri:Habitat} \rightarrow \texttt{Sector} \rightarrow \texttt{agri:Agriculture}$ while for forestry, the aggregation path is $\texttt{agri:Sector}\rightarrow \texttt{agri:Agriculture}$.  The time dimension has one hierarchy \texttt{agri:TimeHierarchy}, and the aggregation path of the hierarchy is $\texttt{agri:Time}\rightarrow \texttt{agri:All}$.

Each level is defined by a set of attributes that describe the characteristics of its members. A level has one or more identifiers used as the primary key and foreign key that uniquely identify the member of a level, and each identifier consists of one or more attributes. Such as, in Figure \ref{fig:schemaDiag1}, \texttt{agri:Agriculture}, \texttt{agri:Sector}, \texttt{agri:Category}, \texttt{agri:Habitat}, \texttt{agri:Product}, \texttt{agri:Time}, \texttt{agri:District}, and \texttt{agri:Division} levels' unique identifiers are \texttt{agri:agricultureId}, \texttt{agri:sect-}\\ \texttt{orId}, \texttt{agri:categoryId}, \texttt{agri:habiatId}, \texttt{agri:productId}, \texttt{agri:yearId}, \texttt{agri:districtId}, and \texttt{agri:divisionId}.

\subsection{Methodology of Generating Bangladesh Agricultural Knowledge Graph (BDAKG) }  \label{sec:md}
\begin{figure*}[h!]
\centering
\includegraphics[scale=0.65]{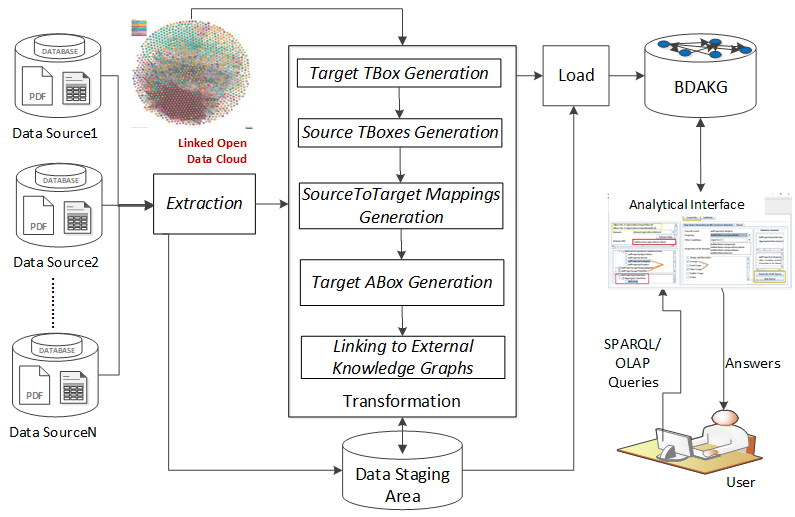}
    \caption{ Overview of the BDAKG generation process.}
    \label{fig:integration}
\end{figure*}

In this section, we delineate the methodology employed for generating the knowledge graph BDAKG.  Figure~\ref{fig:integration} illustrates the various steps involved in the Extraction-Transformation-Load (ETL) process. Initially, the \textit{Extraction} process gathers agriculture related data from diverse sources. Subsequently, the \textit{Transformation} process transforms the extracted data according to the semantics encoded in the target TBox. Within the \textit{Transformation} process, we first implement the target TBox defined in Section~\ref{sec:ttd} through the \textit{Target TBox Generation} process. Following this, the \textit{Source TBoxes Generation} process generates the TBoxes from the data sources. Subsequently, source and target TBoxes are mapped through the \textit{SourceToTarget Mappings Generation} process. Then, the \textit{Target ABox Generation} process produces assertions in accordance with the target TBox. Finally, the resources of target ABox and TBox are linked with external KGs available in the Linked Open Data (LOD) cloud. All interim results are stored in a \textit{Data Staging Area}, a temporary repository that retains data for cleansing, transforming, and future utilization. This staging area safeguards against potential loss of extracted or transformed data in case of loading process failure. Ultimately, the Load operation transfers the knowledge graph to a triple store accessible to users via either an interactive analytical interface or a SPARQL endpoint. The subsequent sections provide detailed insights into each component. 

\subsubsection{Extraction}

Agricultural data for Bangladesh is publicly accessible from various online sources as outlined in Table~\ref{tab:soucedataset}, and it exists in different formats such as PDF, CSV, and Text, etc. The data is extracted and formatted to align with the target schema. Initially, all PDF files are converted to CSV files using a PDF processing tool available at \url{https://www.ilovepdf.com/}. Subsequently, microdata is obtained by eliminating aggregated information, and district, fiscal year, and crop names are replaced with their respective codes. The unique identifier for each crop's information is generated by concatenating districtCode, yearCode, and cropCode. For instance, Table~\ref{table:table1} illustrates the area and production of bananas by district and time, sourced from BBS yearbooks. After extraction and cleansing, the processed data is organized according to the fact table \texttt{agri:Production}, as demonstrated in Table~\ref{minor crops}. Table~\ref{minor crops} presents a segment of the estimated national production and acre-wise for various minor grains from 2017–18 to 2019–20. Only production and acre-wise data for minor crops (barley, bazra, laushak, etc.) are considered, categorized by district and fiscal year. 
As an example of a level dataset, we present selected segments of an \texttt{agri:District} level dataset in Table \ref{table:district}, where the districts are encoded following the conventions of \href{https://www.geonames.org/BD/administrative-division-bangladesh.html}{geonames.org}.

\begin{table*}[h]
\centering
\caption{Area and production of banana by district, 2017-2018 to 2019-20}
\label{table:table1}
\begin{tabular}{|rl|rr|rr|rr|}
\hline
\multicolumn{2}{|l|}{\multirow{2}{*}{District/ Division}}     & \multicolumn{2}{c|}{2017-18}                                                                                                                                          & \multicolumn{2}{c|}{2018-19}                                                                                                                                          & \multicolumn{2}{c|}{2019-20}                                                                                                                                          \\ \cline{3-8} 
\multicolumn{2}{|l|}{}                                        & \multicolumn{1}{c|}{\begin{tabular}[c]{@{}c@{}}Area\\      (acre)\end{tabular}} & \multicolumn{1}{c|}{\begin{tabular}[c]{@{}c@{}}Production\\      (MT)\end{tabular}} & \multicolumn{1}{c|}{\begin{tabular}[c]{@{}c@{}}Area\\      (acre)\end{tabular}} & \multicolumn{1}{c|}{\begin{tabular}[c]{@{}c@{}}Production\\      (MT)\end{tabular}} & \multicolumn{1}{c|}{\begin{tabular}[c]{@{}c@{}}Area\\      (acre)\end{tabular}} & \multicolumn{1}{c|}{\begin{tabular}[c]{@{}c@{}}Production\\      (MT)\end{tabular}} \\ \hline
\multicolumn{1}{|r|}{1}          & Barguna                    & \multicolumn{1}{r|}{331}                                                        & 1132                                                                                & \multicolumn{1}{r|}{338}                                                        & 475                                                                                 & \multicolumn{1}{r|}{347}                                                        & 1580                                                                                \\ \hline
\multicolumn{1}{|r|}{2}          & Barishal                   & \multicolumn{1}{r|}{1668}                                                       & 3219                                                                                & \multicolumn{1}{r|}{1684}                                                       & 3401                                                                                & \multicolumn{1}{r|}{1750}                                                       & 5500                                                                                \\ \hline
\multicolumn{1}{|r|}{3}          & Bhola                      & \multicolumn{1}{r|}{513}                                                        & 1178                                                                                & \multicolumn{1}{r|}{520}                                                        & 1180                                                                                & \multicolumn{1}{r|}{536}                                                        & 1879                                                                                \\ \hline
\multicolumn{1}{|r|}{4}          & Jhallokati                 & \multicolumn{1}{r|}{2824}                                                       & 7461                                                                                & \multicolumn{1}{r|}{2830}                                                       & 7470                                                                                & \multicolumn{1}{r|}{2902}                                                       & 8324                                                                                \\ \hline
\multicolumn{1}{|r|}{5}          & Patuakhali                 & \multicolumn{1}{r|}{764}                                                        & 3343                                                                                & \multicolumn{1}{r|}{765}                                                        & 3345                                                                                & \multicolumn{1}{r|}{554}                                                        & 2717                                                                                \\ \hline
\multicolumn{1}{|r|}{6}          & Pirojpur                   & \multicolumn{1}{r|}{3240}                                                       & 14034                                                                               & \multicolumn{1}{r|}{3280}                                                       & 13386                                                                               & \multicolumn{1}{r|}{2768}                                                       & 13390                                                                               \\ \hline
\multicolumn{1}{|r|}{\textbf{1}} & \textbf{Barishal Division} & \multicolumn{1}{r|}{\textbf{9340}}                                              & \textbf{30367}                                                                      & \multicolumn{1}{r|}{\textbf{9417}}                                              & \textbf{29257}                                                                      & \multicolumn{1}{r|}{\textbf{8857}}                                              & \textbf{33390}                                                                      \\ \hline
\end{tabular}
\end{table*}

\begin{table*}[h!]
    \caption{Representation of Production cuboid crops after processing of data enlisted in Table~\ref{table:table1}} 
   \label{minor crops}
   \small
   \centering
   \begin{tabular}{p{3cm}|p{1.5cm}|p{1.5cm}|p{1.5cm}|p{1.5cm}|p{1.5cm}}
   \toprule\toprule
observationId  & cropsId & districtId & yearId & area & production \\ 
 \midrule
A0101921004201718 & A010192 & 1004 & 201718 & 331 & 1132 \\ 
\midrule
 A0101921006201718 & A010192 & 1006 & 201718 & 1668 & 3219 \\  
\midrule
A0101921004201819 & A010192 & 1004 & 201819 & 338 & 475\\ 
\midrule
A0101921006201819 & A010192 & 1006 & 201819 & 1664& 6401 \\ 
\midrule
 A0101921004201920 & A010192 & 1004 & 201920 & 347 & 1580 \\  
\midrule
 A0101921006201920 & A010192 & 1006 & 201920 & 1750 & 5500 \\  
\bottomrule
\end{tabular}
    
\end{table*}

\begin{table}[h!]
    \caption{Representation of District Level Dataset} 
   \label{table:district}
   \small
   \centering
   \begin{tabular}{p{2cm}|p{2cm}|p{2cm}}
   \toprule\toprule
districtId &	districtName &	inDivision \\ \midrule
1004&	BARGUNA	&10\\ \midrule
1006&	BARISAL&	10 \\ \midrule
1009&	BHOLA&	10 \\ \midrule

\bottomrule

\end{tabular}

\end{table}

\subsubsection{Transformation}
After completing the data extraction and preprocessing processes, we transform all the data into a semantic version to enrich our BDAKG. The \textit{Transformation} process includes following steps, namely

 \begin{itemize}
     \item[-] Target TBox Generation
     \item[-] Source TBox Generation
     \item[-] SourceToTarget mappings Generation
     \item[-] Target ABox Generation
 \end{itemize}
\paragraph{\textbf{Target TBox Generation}}
The aim of this step is to express the specified target TBox outlined in Section~\ref{sec:ttd} utilizing the constructs of RDFS, OWL, and QB4OLAP (as defined in Section~\ref{sec:mmqv}) in conjunction with the RDF model. QB4OLAP constructs enhance OWL classes and RDF properties with MD semantics. Users have the option to create a TBox with MD semantics either manually or by employing tools like Protege~~\cite{musen2015protege}, $SETL_{BI}$~\cite{deb2020setlbi}. Listing~\ref{lst:agricultureTarget} shows a portion of our BDAKG TBox annoted with QB4OLAP constructs.
\lstset{style=mystyle}


\begin{lstlisting}[language=Octave,caption= QB4OLAP representation of BDAKG TBox.,label=lst:agricultureTarget]
@prefix qb: <http://purl.org/linked-data/cube#>.
@prefix xsd: <http://www.w3.org/2001/XMLSchema#>.
@prefix qb4o: <http://purl.org/qb4olap/cubes#>.
@prefix agri: <http://bike-csecu.com/datasets/agri/tbox#>.
@prefix rdf:  <http://www.w3.org/1999/02/22-rdf-syntax-ns#>.
@prefix rdfs: <http://www.w3.org/2000/01/rdf-schema#> .
@prefix owl: <http://www.w3.org/2002/07/owl#> .

#DIMENSIONS
agri:ProductDim a qb:DimensionProperty, owl:Class;
qb4o:hasHierarchy agri:productCropsHierarchy, agri:productFisheriesHierarchy;
rdfs:label "It covers the  prodcut, category, sector ,fisheries etc. level."@en.
#HIERARCHIES
agri:productFisheriesHierarchy a qb4o:Hierarchy, owl:Class;
	rdfs:label "Production Hierarchy O2."@en;
	qb4o:inDimension agri:ProductDim;
	qb4o:hasLevel agri:Agriculture, agri:Sector, agri:Habitat.
#LEVELS
agri:Habitat a qb4o:LevelProperty, owl:Class;
	rdfs:label "Fisheries habitatName and description."@en;
	qb4o:hasAttribute agri:habitatId, agri:habitatName, agri:inSector.
agri:Sector a qb4o:LevelProperty, owl:Class;
	rdfs:label "Fisheries sector name."@en;
	qb4o:hasAttribute agri:sectorId, agri:inAgriculture.
#Hierarchy steps
_:hs1 a qb4o:HierarchyStep;
	qb4o:inHierarchy agri:productFisheriesHierarchy;
	qb4o:childLevel agri:Habitat;
	qb4o:parentLevel agri:Sector;
	qb4o:pcCardinality qb4o:OneToMany;
	qb4o:rollup agri:inSector.
#ATTRIBUTES
agri:habitatId a qb4o:LevelAttribute, rdf:Property, owl:DatatypeProperty;
	rdfs:label "Fisheris habitat."@en;
	rdfs:domain agri:Habitat;
	rdfs:range xsd:integer.
agri:inSector a qb4o:LevelAttribute, rdf:Property, owl:ObjectProperty;
	rdfs:doamin agri:Habitat;
	rdfs:range agri:Sector.
#MEASURES
agri:area a qb:MeasureProperty, rdf:Property;
	rdfs:label "Area required for crops production."@en;
	rdfs:range xsd:float.
#CUBES
agri:agricultureFisheriesCuboid a qb:DataStructureDefinition, owl:Class;
	dct:conformsTo <http://purl.org/qb4olap/cubes>;
	qb:component [ qb:measure agri:area; qb4o:aggregateFunction qb4o:avg, qb4o:count, qb4o:sum];
	qb:component [ qb:measure agri:production; qb4o:aggregateFunction qb4o:avg, qb4o:count, qb4o:sum];
	qb:component [ qb4o:level agri:District; qb4o:cardinality qb4o:OneToMany];
	qb:component [ qb4o:level agri:Habitat; qb4o:cardinality qb4o:OneToMany];
	qb:component [ qb4o:level agri:Time; qb4o:cardinality qb4o:OneToMany].
#DATASETS
agri:agricultureFisheriesDataset a qb:DataSet, owl:Class;
qb:structure agri:agricultureFisheriesCuboid.


\end{lstlisting}

In QB4OLAP, dimensions, hierarchies, and levels are defined using \texttt{qb4o:DimensionProperty, qb4o:Hierarchy, qb4o:LevelProperty} as demonstrated at lines 9-24. A dimension can have one or more hierarchies, and line 11 shows that \texttt{agri:ProductDim} has two hierarchies. The properties \texttt{qb4o:inDimension} and its inverse \texttt{qb4o:hasHierarchy} are employed to establish the relationship between a dimension and its hierarchies (lines 11 and 16).A hierarchy consists of levels arranged in a specific order, and the order of levels in a hierarchy is defined using \texttt{qb4o:HierarchyStep}. In the listing, \texttt{agri:productFisheriesHierarchy} comprises three levels (line 17). The hierarchy step, defined at lines 25-31, illustrates that habitats are aggregated in sectors, i.e., $\texttt{agri:Habitat}\rightarrow \texttt{agri:Sector}$ through the roll-up property \texttt{agri:inSector} (line 31). Attributes of a level are defined using \texttt{qb4o:LevelAttribute} (lines 33-39). It's important to note that a level attribute can either be an object property (relating among instances) or a datatype property (relating instances to literal values). For instance, the \texttt{agri:inSector} attribute establishes a relationship between instances of \texttt{agri:Habitat} and \texttt{agri:Sector}. In QB4OLAP, a cube structure is delineated in terms of dimensions and measures, and a cuboid structure is defined in terms of levels and measures (lines 45-51). Both structures are specified using \texttt{qb:DataStructureDefinition}. Measures, representing the numerical data for analysis and insights, are defined using \texttt{qb:MeasureProperty} (lines 41-43). Finally, a dataset is defined using \texttt{qb:Dataset} and is assigned to the defined cube or cuboid structure (lines 53-54).

\paragraph{\textbf{Source TBox Generation}}
After defining BDAKG at the TBox level, we need to populate it from the available sources. To do so, we need to map the target and source constructs at the TBox level. Therefore, it is crucial to derive TBoxes from the existing sources and enhance them with OWL and RDFS constructs. In the \textit{Extraction} phase, data is extracted and cleansed from various sources, and the resulting information is stored in a tabular format.  In this step, for each table, we derive the schema of those tabular data sources and convert them into source TBoxes. A straightforward approach involves considering the table name as an OWL class and the attribute names as OWL datatype or object properties. Additionally, users have the option to utilize R2RML~\cite{world2012r2rml} or direct mapping~\cite{arenas2012direct} vocabularies to perform the extraction and creation of source TBoxes. Listing \ref{lst:level2} show the source TBox of the Habitat dataset, where \texttt{onto:Habitat} is considered as an OWL class and \texttt{onto:habitatId} and \texttt{onto:inSector} are datatype properties. 


\lstset{style=mystyle}

\begin{lstlisting}[language=Octave, caption= Source TBox of Habitat of the Habitat dataset.,label=lst:level2 ]
@prefix onto: <http://bike-csecu.com/datasets/agri/onto/habitat#>.
@prefix owl: <http://www.w3.org/2002/07/owl#>.
@prefix rdfs: <http://www.w3.org/2000/01/rdf-schema#>.
onto:Habitat a owl:Class.
onto:habitatId a owl:DatatypeProperty;
        rdfs:domain onto:Habitat;
        rdfs:range xsd:string.
onto:inSector a  owl:DatatypeProperty;
        rdfs:domain onto:Habitat;
        rdfs:range xsd:string.
\end{lstlisting}


\paragraph{\textbf{SourceToTarget Mappings Generation}}
To map between the soruce and target TBox constructs, we use Source-to-Target Mapping (S2TMAP) vocabulary~\cite{deb2022high}: an OWL-based mapping vocabulary.  In Listing ~\ref{lst:map}, the mapping definitions between BDAKG's TBox and Habitat's TBox are presented, with annotations incorporating S2TMAP constructs.  
\lstset{style=mystyle}
 

\begin{lstlisting}[language=Octave,caption=SourceToTarget mapping definitions between the target TBox \texttt{agri:} and the source TBox \texttt{onto:Habitat}.,label=lst:map]
@prefix map: <http://bike-csecu.com/datasets/vocabulary/s2tmap#>.

#Dataset mapper
map:habitatDataset  a  map:Dataset;
        map:sourceTBox  <http://bike-csecu.com/datasets/agri/onto/habit#>;
        map:targetTBox  <http://bike-csecu.com/datasets/agri/tbox#>.
#Concept mapper
map:Habitat_Habitat  a            map:ConceptMapper;
        map:dataset               map:habitatDataset;
        map:iriValue              onto:habitatId;
        map:iriValueType          map:SourceAttribute;
        map:matchedInstances      "All";
        map:sourceConcept         onto:Habitat;
        map:targetCommonProperty  onto:habitatId;
        map:targetConcept         agri:Habitat.
#Property mapper
map:PropertyMapper_01_habitatId_habitatId
        a                       map:PropertyMapper;
        map:ConceptMapping       map:Habitat_Habitat;
        map:sourceProperty      onto:habitatId;
        map:sourcePropertyType  map:SourceProperty;
        map:targetProperty      agri:habitatId.

\end{lstlisting}

Within S2TMAP, it is possible to specify a property-level mapping nested within a concept-level mapping, and this, in turn, is defined within a mapping dataset. A mapping dataset is established through the use of \texttt{map:Dataset}, capturing the addresses of the source and target TBoxes (lines 4-6 in Listing~\ref{lst:map}). A concept-mapping delineates the correspondence between a source and a target concept (lines 8-15). The linkage between a concept-mapping and its mapping dataset is established via the \texttt{map:dataset} property. The \texttt{map:iriValue} and \texttt{map:iriValueType} properties signify that the values of \texttt{onto:habitatId} will be utilized to generate IRIs for the members of \texttt{agri:Habitat}. The "All" value of \texttt{map:matchedInstances} indicates that all source instances are mapped. A property-mapping is employed for mapping at the property level (lines 17-21). The association between a property-mapping and its corresponding concept-mapping is established via \texttt{map:conceptMapping}. The target property of the property-mapping is specified using \texttt{map:targetProperty}, and this target property can be linked to either a source property or an expression. In this particular instance, the target property \texttt{agri:habitatId} is mapped to the source property \texttt{onto:habitatId}.

\paragraph{\textbf{Target ABox Generation}} 
Taking the target TBox, source datasets (extracted and cleansed ones), source-to-target mapping definitions as inputs, the \textit{Target ABox Generation} process creates the ABox of BDAKG from the source datasets according to the semantics encoded in the TBox of BDAKG. Within QB4OLAP, dimensional data is stored physically in levels. Each level member is characterized by a distinct IRI and is semantically associated with its corresponding level attributes and roll-up properties. For example, Listing~\ref{lst:level} shows a member of \texttt{agri:Product}. It is defined as a \texttt{qb4o:LevelMember} and semantically enriched with values of its linked properties. Note that, with the value of \texttt{agri:inCategory} (a roll-up property), the product is connected with its  category, an level member of \texttt{agri:Category}.

\lstset{style=mystyle}


\begin{lstlisting}[language=Octave,caption= A level member of \texttt{agri:Product}.,label=lst:level]
@prefix product: <http://bike-csecu.com/datasets/agri/abox/product#>.
@prefix category: <http://bike-csecu.com/datasets/agri/abox/category#>.
product:A010189 a qb4o:LevelMember;
        qb4o:memberOf agri:Product;
        agri:cropsCode "189";
        agri:cropsId "A010189";
        agri:cropsName "Cabbage";
        agri:harvestTime "Early January to Early March";
        agri:inCategory category:A01010;
        agri:sawingTime "Late October to  Mid November";
        agri:scientificName "Brassica oleracea var. capitata".
\end{lstlisting}

\lstset{style=mystyle}


\begin{lstlisting}[language=Octave,caption=An observation of the \texttt{agri:Production} cuboid.,label=lst:fact]
@prefix production: <http://bike-csecu.com/datasets/agri/abox/production#>.
@prefix district: <http://bike-csecu.com/datasets/agri/abox/district#>.
@prefix product: <http://bike-csecu.com/datasets/agri/abox/product#>.
@prefix time: <http://bike-csecu.com/datasets/agri/abox/time#>.
#observation
production:A0101921004201718 a qb:Observation;
        qb:dataSet agri:agricultureDataset;
        agri:District district:1004;
        agri:Product product:A010192;
        agri:Time time:201718;
        agri:area "331";
        agri:production "1132".
\end{lstlisting}

To represent a fact, QB4OLAP uses an observation (an instance of \texttt{qb:Observation}) (line 6 in Listing~\ref{lst:fact}). A fact is defined by a unique IRI and is semantically enriched through a combination of multiple members from various levels, incorporating values for different measure properties. Note that, the dataset of the fact is \texttt{agri:agricultureDataset}, and its cuboid structure comprises levels such as \texttt{agri:District, agri:Product, agri:Time}, along with measures including \texttt{agri:area} and \texttt{agri:production}.

\paragraph{\textbf{Linking to external knowledge graphs}}\label{sec:linking} Linked open datasets incorporate references to comparable elements found in external datasets, facilitating the sharing and reuse of existing knowledge. This practice aids in steering clear of redundant data inclusion, thereby contributing to the preservation of scalability. Linking can extend to both concepts in the target TBox and instances at the level in the target ABox. BDAKG is linked to  Wikidata~\cite{vrandevcic2014wikidata}, \href{https://www.geonames.org/ontology/documentation.html} {Geonames}, Exiobase~\cite{ghose2022core}, and CropOntology \cite{matteis2013crop}.  This is
achieved using the OWL property \texttt{owl:sameAs}. For example, the triple <\texttt{ agri:District owl:sameAs wiki:Q152732}.> indicates \texttt{ agri:District} is connected to  \texttt{ wiki:Q152732} . We utilize OpenRefine~\cite{verborgh2013using} to create connections between internal and external resources.

\subsection{Load}
BDAKG is expressed through RDF triples, and we utilize Turtle~ \cite{TurtleTe14:online} as the RDF serialization format. The \textit{Load} process involves loading BDAKG into either the triple store, Virtuoso, or as a dump to a local file.


\section{Results and Discussion}\label{sec:exp}

In this section we describe the overview of the produced knowledge graph: BDAKG and assess BDAKG in terms of three aspects: 1) ETL performance, i.e., the time required to run the ETL process; 2) quality of BDAKG, including its compatibility with OLAP operations; and 3) its preparedness for exploratory analytics, gauged by its ability to navigate for exploration and insight generation. The experiments are conducted on a laptop equipped with an Intel Core(TM) i7-4600U processor running at 2.10 GHz, 8 GB of RAM, and operating on Windows 10.

\subsection{Description of the produced knowledge graph BDAKG}\label{sec:description} This section provides an overview of BDAKG, covering its dimensions, factual aspects, external links, and accessibility. 
\subsubsection{Dimension overview} Table~\ref{tab:dimensionOverview} outlines the dimensions of BDAKG, the levels contained in each dimension, number of level attributes, members, external links, and RDF triples. In total, BDAKG has 9 levels, 33 level attributes, 272 level members, 363 external links and 1738 RDF triples. All levels are linked to Wikidata~\cite{vrandevcic2014wikidata}. In addition, districts and divisions are connected to \href{https://www.geonames.org/ontology/documentation.html} {Geonames} and products are linked to Exiobase~\cite{ghose2022core}: a semantically annotated global environmentally extended multi-regional input-output database. Exiobase is particularly useful for life cycle assessment (LCA) and environmental impact assessment studies. 

\begin{table}[h!]
\centering
\caption{Overview of dimensions in BDAKG.}
\label{tab:dimensionOverview}
\resizebox{\textwidth}{!}{%
\begin{tabular}{|l|l|r|r|r|r|}
\hline
\textbf{Dimension}                 & \textbf{Level}   & \textbf{\begin{tabular}[c]{@{}r@{}}Number of \\ attributes\end{tabular}} & \textbf{\begin{tabular}[c]{@{}r@{}}Number of\\  instances\end{tabular}} & \textbf{\begin{tabular}[c]{@{}r@{}}Number of\\  external links\end{tabular}} & \textbf{\begin{tabular}[c]{@{}r@{}}Number of\\  RDF tiples\end{tabular}} \\ \hline
\multirow{3}{*}{\texttt{agri:GeographyDim}} & \texttt{agri:District}    & 3                                                                        & 64                                                                      & 128                                                                          & 448                                                                      \\ \cline{2-6} 
                                   & \texttt{agri:Division}    & 3                                                                        & 7                                                                       & 14                                                                           & 49                                                                       \\ \cline{2-6} 
                                   & \texttt{agri:All}         & 3                                                                        & 1                                                                       & 1                                                                            & 6                                                                        \\ \hline
\multirow{5}{*}{\texttt{agri:ProductDim}}   & \texttt{agri:Product}     & 3                                                                        & 114                                                                     & 134                                                                          & 704                                                                      \\ \cline{2-6} 
                                   & \texttt{agri:Category}    & 3                                                                        & 15                                                                      & 15                                                                           & 90                                                                       \\ \cline{2-6} 
                                   & \texttt{agri:Habitat}     & 3                                                                        & 14                                                                      & 14                                                                           & 84                                                                       \\ \cline{2-6} 
                                   & \texttt{agri:Sector}      & 6                                                                        & 4                                                                       & 4                                                                            & 36                                                                       \\ \cline{2-6} 
                                   & \texttt{agri:Agriculture} & 6                                                                        & 1                                                                       & 1                                                                            & 9                                                                        \\ \hline
\texttt{agri:TimeDim}                       & \texttt{agri:Time}        & 3                                                                        & 52                                                                      & 52                                                                           & 312                                                                      \\ \hline
\textbf{Total}                     & \textbf{9}                 & \textbf{33}                                                              & \textbf{272}                                                            & \textbf{363}                                                                 & \textbf{1738}                                                            \\ \hline
\end{tabular}%
}
\end{table}

\subsubsection{Fact overview} 
Within BDAKG, information is organized into three distinct cuboids to capture facts. Table~\ref{tab:fact_overview} displays the ABox size, the count of observations, and the number of RDF triples for each of these cuboids. In summary, the total size of cuboids, number of observations, the number of triples amount to 42 MB, 55048, and 379645 respectively.  

\begin{table*}[h!]
\centering
\caption{Overview of the size metrics of the cuboids. }
\label{tab:fact_overview}
\begin{tabular}{|l|r|r|r|}
\hline
\textbf{Cuboid}        &  \textbf{ABox Size}           & \textbf{Number of observations} & \textbf{Number of RDF triples} \\ \hline

\texttt{agri:Production} &  36.7 MB & 49459 &   338355      \\ \hline
\texttt{agri:Fisheries}      &5 MB & 5205   &    39004    \\ \hline
\texttt{agri:Forestry}       & 300 KB    & 384    &      2286  \\ \hline
\textbf{Total}              & \textbf{42 MB}    & \textbf{55048}    &      \textbf{379645}  \\ \hline
\end{tabular}
\end{table*}

\subsubsection{Availability}
\label{sec:availability}

The dump files for our knowledge graph are located at \url{http://bike-csecu.com/datasets/agri/}, and the knowledge graph itself is stored in the OpenLink Virtuoso Triplestore. Users have the option to remotely access the knowledge graph via the SPARQL endpoint at \url{http://bike-csecu.com:8893/sparql/}. They can formulate their own SPARQL queries based on their specific requirements to obtain the desired answers.
To verify the correctness and OLAP-compatibility of the knowledge graph, we have formulated a set of competency questions in Table~\ref{tab:competencyQuestions}. These competency questions have been translated into equivalent SPARQL queries to retrieve answers from the knowledge graph. Access to this set of competency questions is available through a user interface provided at \url{https://bdakg.netlify.app}. Users can pose these questions to the repository and receive answers.
Additionally, we offer an interactive OLAP interface, accessible at \url{https://github.com/bi-setl/SETL}. This interface enables users to create OLAP queries using graphical user interface components and retrieve answers by submitting queries to the relevant graphs. Details about the OLAP interface are provided in Section~\ref{sec:olap}.

\subsection{ETL Performance}
\begin{figure*}[h!]
\centering
\includegraphics[scale=.42]{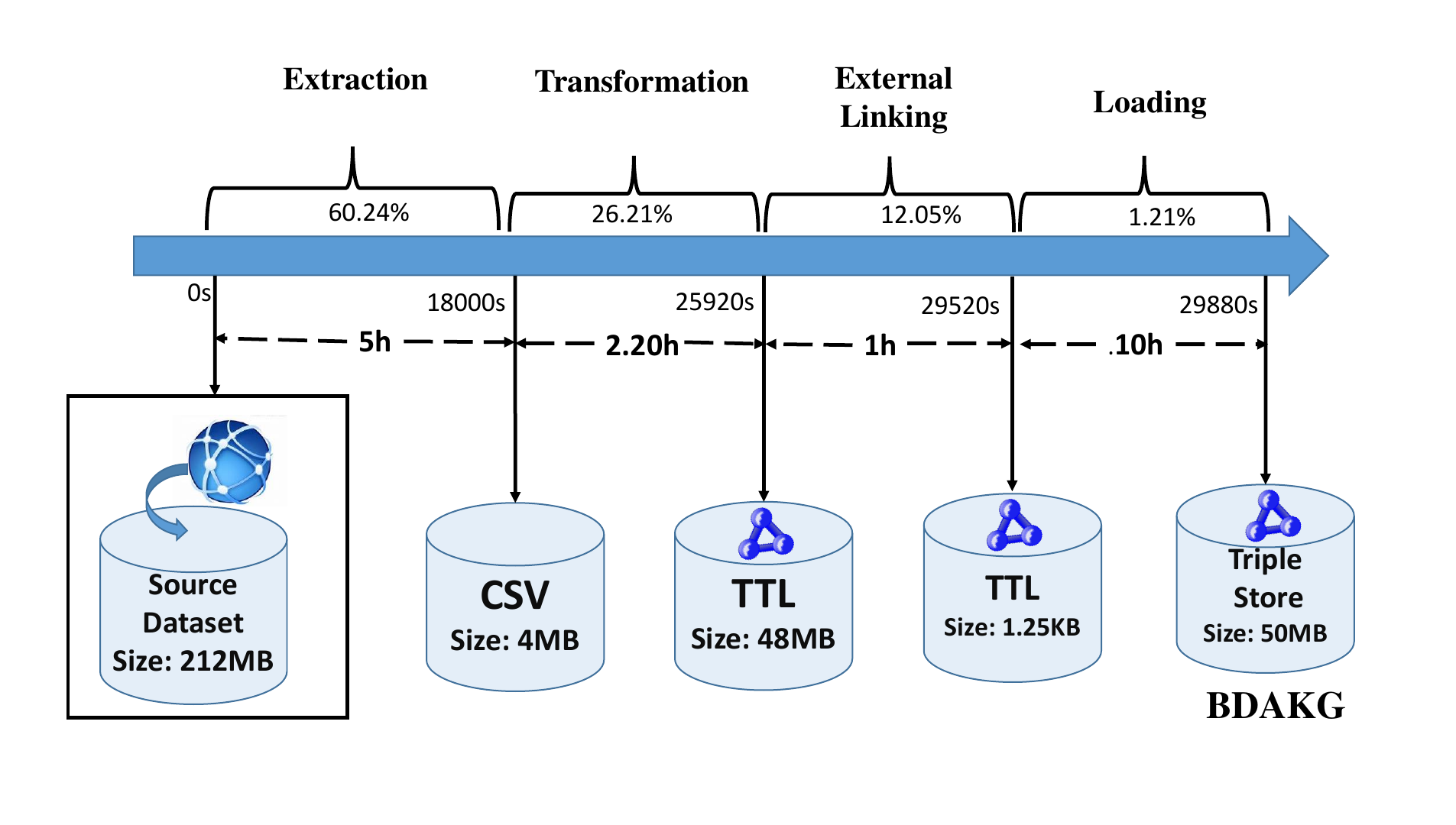}
    \caption{ Time for the ETL process to create BDAKG. Here, h and s indicate hour and second respectively.}
    \label{fig:ETLPermo}
\end{figure*}

This section focuses on the ETL process's time required for the different stages of creating BDAKG. The duration and percentage of total time that is used on each step of ETL operations are shown in Figure \ref{fig:ETLPermo}. The \textit{Extraction} phase takes most of the overall time (60.24\%). For the following reasons, this phase requires the most time in comparison to other phases: First, the Bangladesh agriculture data available at BBS is provided in a read-only format; therefore, we need to do extra pre-processing, cleansing, and formatting tasks to extract the attribute information from them and keep it into the tabular format.  
We utilize the $SETL_{BI}$ tool to generate TBoxes, source-to-target mapping definitions, and ABox. The entire \textit{Transformation} phase consumes a total of 2.2 hours. Note that the TBox and mapping definition creation processes involve user intervention, leading to variable time-frames depending on user expertise. Additionally, we employ OpenRefine~\cite{verborgh2013using} to establish links between internal and external resources, and, similar to the prior processes, the elapsed time may differ based on user proficiency due to the required user interaction.

\subsection{Quality of BDAKG}
We evaluate the quality of BDAKG using various standard metrics, including completeness, timeliness, granularity, OLAP compatibility, and correctness. It is noteworthy that our knowledge graph excels in the majority of these metrics, attributed to its formation from pre-existing, well-organized, and dependable non-semantic data.

\subsubsection{Completeness} Completeness \cite{batini2016data} describes the extent to which a given dataset contains all necessary data. We demonstrate completeness in BDAKG across three aspects: schema completeness, property completeness, and linkability completeness.

\textbf{Schema Completeness:} Our agriculture constellation schema Figure \ref{fig:schemaDiag1} refers to the extent to which a data schema or data model effectively captures all the necessary information, attributes, entities, relationships, and constraints required to represent a real-world agriculture dataset. Such as, when a new crop is produced in Bangladesh, we can categorize it according to its name, crop ID, scientific name, harvest time, and sowing time, along with the characteristics of the crop.\par
\textbf{Property Completeness} 
The degree of missing values for a specific property is evaluated through its property completeness, as defined in \cite{kg-book}. For instance, our BDAKG exhibits a property completeness issue in cases such as the absence of a scientific name for instances like Onion, which should ideally have the value \textit{Allium cepa}. This deficiency is also referred to as column completeness \cite{pipino2002data}. In specific instances, particularly for products (level members of \texttt{agri:Product}), we encounter challenges in determining sowing and harvesting times. The calculation of the property completeness for our knowledge graph is determined using Equation \ref{eq:prop_complete}, where P denotes the percentage of property completeness \cite{pipino2002data}. Our analysis reveals that the property completeness score for \texttt{agri:Product} is 93.97\%. 

\begin{equation}
    P =[1- \left(\frac{\text{Number of incomplete items}}{\text{Total number of items}}\right)] \times 100
    \label{eq:prop_complete}
\end{equation}

\textbf{Population Completeness} This evaluates~\cite{issa2021knowledge} how effectively BDAKG represents the real-world objects. For instance, there are four sectors of agriculture in BDAKG, and in reality, Bangladesh's agriculture sectors are also classified into four categories: crop, fisheries, forestry, and livestock. BDAKG contains all products, firsheries, and forestry data found in the available open data.

\textbf{Linkability Completeness:} 
In Section~\ref{sec:linking}, it is explained that resources within BDAKG are interconnected with Wikidata, Cropontology, Geonames, RDF, OWL, and QB4OLAP at both TBox and ABox levels. Consequently, this connectivity facilitates the exploration of other knowledge graphs for enhanced insights, as detailed in Section~\ref{section:analytics}.

\subsubsection{Timeliness and granularity}
The BDAKG dataset is current, encompassing the most recent data for Fisheries (2019–2020 fiscal year), Forestry (2017–2018 fiscal year), and Agriculture (2019–2020 fiscal year). To preserve granularity, only district-level data on crops, fisheries, and forestry is retained in the BDAKG dataset. All aggregate data at the division and country levels are excluded. This modeling approach ensures that the BDAKG dataset retains valuable information for informed decision-making.

\subsection{Data FAIRness}
BDAKG adheres to the four key constructs of the FAIR principles as follows: 
\begin{itemize}
\item[-] Findability: Each resource in BDAKG, whether at the TBox or ABox level, is uniquely identified by an Internationalized Resource Identifier (IRI). For instance, the knowledge graph's dump files are accessible at \url{http://bike-csecu.com/datasets/agri/}. Additionally, BDAKG's TBox and ABox can be accessed through \url{https://bike-csecu.com/datasets/agri/abox/abox.ttl} and \url{https://bike-csecu.com/datasets/agri/tbox/tbox.ttl}, respectively.
\item[-] Accessibility: Users can readily access and download BDAKG from the specified repository. Data retrieval is also possible through the SPARQL endpoint, available at \url{http://bike-csecu.com:8893/sparql/}.
\item[-] Interoperability: BDAKG employs standard vocabularies such as RDF, RDFS, OWL, and QB4OLAP. This adherence to standards ensures that BDAKG can be seamlessly integrated and utilized with other datasets and tools.
\item[-] Reusability: BDAKG includes comprehensive metadata and licensing information, promoting its reusability. Users can reproduce the dataset following the outlined steps in this paper. Therefore, individuals can easily reuse the datasets or resources within BDAKG for their specific needs.

\end{itemize}

\subsubsection{OLAP compatibility}\label{sec:olap}

Our BDAKG knowledge graph adheres to the formal semantics and constraints defined at the TBox level. The TBox is annotated with multidimensional semantics using QB4OLAP, and the ABox, generated through the KG generation process, aligns with the semantics specified in the TBox. In this section, we assess the OLAP compatibility of BDAKG. The assessment involves the utilization of $SETL_{BI}$'s \textit{OLAP Layer}~\cite{deb2020setlbi} to verify BDAKG's OLAP compatibility. We load BDAKG into the \textit{OLAP layer}, formulate and execute OLAP queries, and observe error-free outputs. This affirms the successful enhancement of the knowledge graph for business analytics.
\begin{figure*}[h!]
\centering
\includegraphics[scale=0.5]{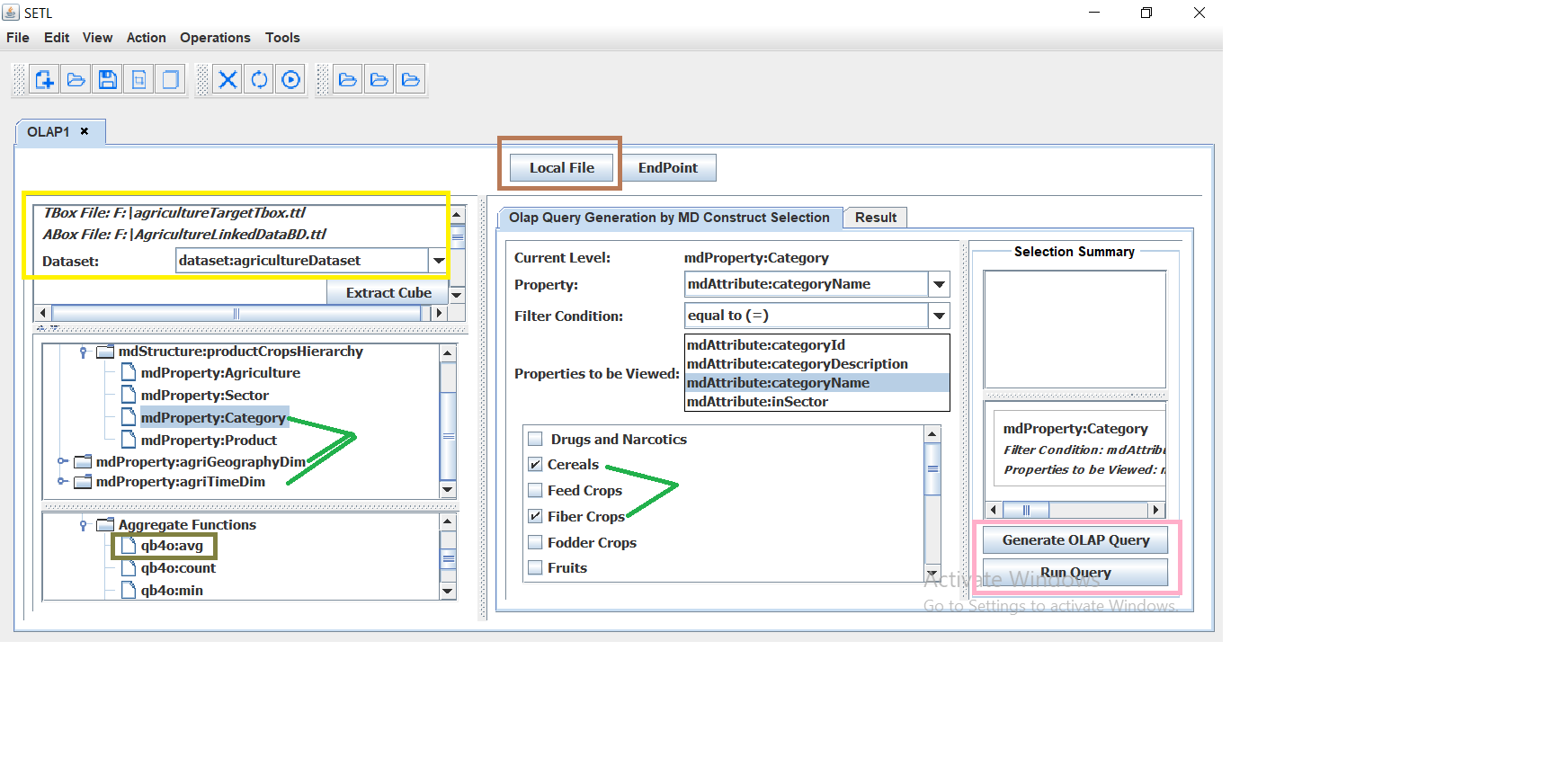}
    \caption{Enabling OLAP operations over BDAKG throgh $SETL_{BI}'s \;\;OLAP Layer$.}
    \label{olaptool}
\end{figure*}

\begin{lstlisting}[language=Python, caption=Query example,label=Query]
PREFIX qb: <http://purl.org/linked-data/cube#>
PREFIX qb4o: <http://purl.org/qb4olap/cubes#>
PREFIX skos: <http://www.w3.org/2004/02/skos/core#>
SELECT ?agriProductDim_categoryName ?agriGeographyDim_divisionName ?agriTimeDim_yearName (AVG(<http://www.w3.org/2001/XMLSchema#float>(?m1)) as ?area_avg) 
(AVG(<http://www.w3.org/2001/XMLSchema#float>(?m2)) as ?production_avg) 
WHERE {
?o a qb:Observation .
?o qb:dataSet <http://bike-csecu.com/datasets/agri/abox/data#agricultureDataset> .
?o <http://bike-csecu.com/datasets/agri/abox/mdProperty#area> ?m1 .
?o <http://bike-csecu.com/datasets/agri/abox/mdProperty#production> ?m2 .
?o <http://bike-csecu.com/datasets/agri/abox/mdProperty#Product> ?agriProductDim_Product .
?agriProductDim_Product qb4o:memberOf <http://bike-csecu.com/datasets/agri/abox/mdProperty#Product> .
?agriProductDim_Product <http://bike-csecu.com/datasets/agri/abox/mdAttribute#inCategory> ?agriProductDim_Category .
?agriProductDim_Category qb4o:memberOf <http://bike-csecu.com/datasets/agri/abox/mdProperty#Category> .
?agriProductDim_Category <http://bike-csecu.com/datasets/agri/abox/mdAttribute#categoryName> ?Category_categoryName .
?agriProductDim_Category <http://bike-csecu.com/datasets/agri/abox/mdAttribute#categoryName> ?agriProductDim_categoryName .
?o <http://bike-csecu.com/datasets/agri/abox/mdProperty#District> ?agriGeographyDim_District .
?agriGeographyDim_District qb4o:memberOf <http://bike-csecu.com/datasets/agri/abox/mdProperty#District> .
?agriGeographyDim_District <http://bike-csecu.com/datasets/agri/abox/mdAttribute#inDivision> ?agriGeographyDim_Division .
?agriGeographyDim_Division qb4o:memberOf <http://bike-csecu.com/datasets/agri/abox/mdProperty#Division> .
?agriGeographyDim_Division <http://bike-csecu.com/datasets/agri/abox/mdAttribute#divisionName> ?agriGeographyDim_divisionName .
?o <http://bike-csecu.com/datasets/agri/abox/mdProperty#Time> ?agriTimeDim_Time .
?agriTimeDim_Time <http://bike-csecu.com/datasets/agri/abox/mdAttribute#yearName> ?Time_yearName .
?agriTimeDim_Time qb4o:memberOf <http://bike-csecu.com/datasets/agri/abox/mdProperty#Time> .
?agriTimeDim_Time <http://bike-csecu.com/datasets/agri/abox/mdAttribute#yearName> ?agriTimeDim_yearName .
FILTER ((REGEX (?Category_categoryName, "Cereals", "i") || REGEX (?Category_categoryName, "Fiber Crops", "i")) && REGEX (?Time_yearName, "2018-19", "i"))}
GROUP BY ?agriProductDim_categoryName ?agriGeographyDim_divisionName ?agriTimeDim_yearName ?area_avg
ORDER BY ?agriProductDim_categoryName ?agriGeographyDim_divisionName ?agriTimeDim_yearName ?area_avg
\end{lstlisting}

\begin{figure*}[h!]

\centering
\includegraphics[scale=0.5]{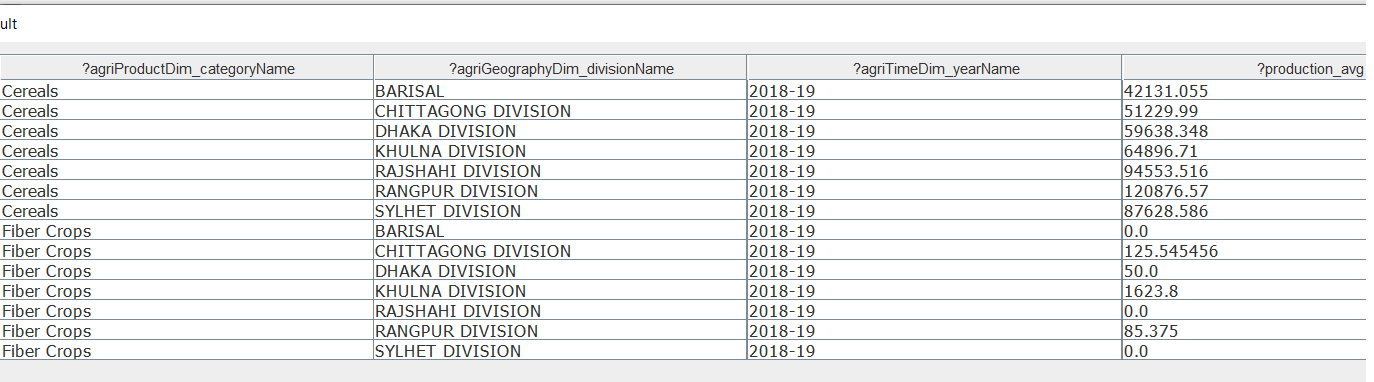}
    \caption{ Query Result.}
   \label{QueryResult}
\end{figure*}

Figure \ref{olaptool} displays the interactive interface of the \textit{OLAP layer}, indicating the successful loading of the TBox and ABox of BDKAG, as highlighted in the upper-left corner (enclosed by the yellow rectangle). Subsequently, within the visualization panel, we specify the levels, measures, and aggregation functions to formulate the desired query. Finally, by activating the buttons labeled within a deep-yellow rectangle (Run Query and Generate OLAP Query), we can examine the output and view the corresponding SPARQL query code, respectively. The generated query is provided in Listing \ref{Query}, and the query result is presented in Figure \ref{QueryResult}. This SPARQL query is designed to fetch and aggregate data related to \texttt{agri:Production} by selecting \texttt{agri:Category}, \texttt{agri:Division}, and \texttt{agri:Year} from \texttt{agri:ProductDim, agri:GeographyDim, agri:TimeDim}, respectively. This demonstration affirms the OLAP compliance of BDAKG. On top of that, BDAKG also enables inter-cube query. An example of an inter-cube query illustrating district-wise crops and fisheries production can be found at \url{https://bdakg.netlify.app}.

\begin{table*}[]
\centering
\caption{A set of competency queries and correctness of BDAKG.}
\label{tab:competencyQuestions}
\begin{adjustbox}{angle=90,width=.75\textwidth}
\resizebox{\textwidth}{!}{%
\begin{tabular}{|l|l|l|l|l|l|c|}
\hline
\textbf{\begin{tabular}[c]{@{}l@{}}No of \\ Query\end{tabular}} & \textbf{\begin{tabular}[c]{@{}l@{}}Types of\\ Query\end{tabular}}              & \textbf{\begin{tabular}[c]{@{}l@{}}Target\\ Cuboid\end{tabular}}             & \textbf{Query Text}                                                                                                                                                               & \textbf{\begin{tabular}[c]{@{}l@{}}Answer from \\ original dataset\end{tabular}}                 & \textbf{\begin{tabular}[c]{@{}l@{}}Answer from \\ BDAKG\end{tabular}}                            & \textbf{\begin{tabular}[c]{@{}c@{}}Is BDAKG\\ correct?\end{tabular}} \\ \hline
Q1                                                              &                                                                                & \texttt{agri:Production}                                                              & \begin{tabular}[c]{@{}l@{}}What is the total production  per acre for each crop category in Bangladesh division-wise?\end{tabular} & \begin{tabular}[c]{@{}l@{}}Not Applicable \end{tabular}                       & \begin{tabular}[c]{@{}l@{}}Can Give Ans\end{tabular}                       & Y                                                                    \\ \cline{1-1} \cline{3-7} 
Q2                                                              &                                                                                & \texttt{agri:Forestry}                                                                & What is the total forest area in Bangladesh divion-wise?                                                                                                                & Not Applicable                                                                                   & Can Give Ans                                                                                   & Y                                                                    \\ \cline{1-1} \cline{3-7} 
Q3                                                              & \multirow{-3}{*}{Roll-up}                                                      & \texttt{agri:Fisheries}                                                               & How many metric tons of total fish production in Bangladesh in division-wise?                                                                                                           & Not Applicable                                                                                       & Can Give Ans                                                                                  & Y                                                                    \\ \hline
Q4                                                              &                                                                                & \texttt{agri:Production}                                                              & \begin{tabular}[c]{@{}l@{}}What is the total production  per acre for each crop category \\in Bangladesh district-wise?\end{tabular}                                                          & \begin{tabular}[c]{@{}l@{}}Not Applicable\end{tabular}                   & \begin{tabular}[c]{@{}l@{}}Can Give Ans\end{tabular}                   & Y                                                                    \\ \cline{1-1} \cline{3-7} 
Q5                                                              &                                                                                & \texttt{agri:Fisheries}                                                               & \begin{tabular}[c]{@{}l@{}}How many metric tons of total fish production in Bangladesh in district-wise?  \end{tabular}                                                           & Not Applicable                                                                                          & Can Give Ans                                                                                       & Y                                                                    \\ \cline{1-1} \cline{3-7} 
Q6                                                              & \multirow{-3}{*}{Drill-down}                                                   & \texttt{agri:Forestry}                                                                & \begin{tabular}[c]{@{}l@{}}What is the total forest area in Bangladesh district-wise?\end{tabular}                                         & Not Applicable                                                                                        &  Can Give Ans                                                                                        & Y                                                                    \\ \hline
Q7                                                              &                                                                                & \texttt{agri:Forestry}                                                                & \begin{tabular}[c]{@{}l@{}}What is the minimum forest area in the Bandarban district ?\end{tabular}                                                       & Not Applicable                                                                                      &   Can Give Ans                                                                                    & Y                                                                    \\ \cline{1-1} \cline{3-7} 
Q8                                                              &                                                                                & \texttt{agri:Production}                                                              & \begin{tabular}[c]{@{}l@{}}What is the maximum M.Ton onion have been produced  in Chadpur district  in \\the respective area?\end{tabular}                                & {\color[HTML]{374151} \begin{tabular}[c]{@{}l@{}}4318 metric tons\\ and 1724 acres\end{tabular}} & {\color[HTML]{374151} \begin{tabular}[c]{@{}l@{}}4318 metric tons\\ and 1724 acres\end{tabular}} & Y                                                                    \\ \cline{1-1} \cline{3-7} 
Q9                                                              & \multirow{-3}{*}{Slice}                                                        & \texttt{agri:Fisheries}                                                               & \begin{tabular}[c]{@{}l@{}}How many average metric tons of fish have been produced in Baor habitat\\ in Barguna district?\end{tabular}                                                        &    Not Applicable                                                                                        & Can Give Ans                                                                                          & Y                                                                    \\ \hline

Q7                                                              &                                                                                & \texttt{agri:Forestry}                                                                & \begin{tabular}[c]{@{}l@{}}What is the average forest area in the Bandarban  district for the years 2016-17?\end{tabular}                                                       & 797541.49 acres                                                                                      & 797541.49  acres                                                                                      & Y                                                                    \\ \cline{1-1} \cline{3-7} 
Q8                                                              &                                                                                & \texttt{agri:Production}                                                              & \begin{tabular}[c]{@{}l@{}}How many metric tons of onion have been produced  in Chadpur district in 2018-19 \\ in the respective area?\end{tabular}                                & {\color[HTML]{374151} \begin{tabular}[c]{@{}l@{}}4308.0 metric tons\\ and 1661.0 acres\end{tabular}} & {\color[HTML]{374151} \begin{tabular}[c]{@{}l@{}}4308.0 metric tons\\ and 1661.0 acres\end{tabular}} & Y                                                                    \\ \cline{1-1} \cline{3-7} 
Q9                                                              & \multirow{-3}{*}{Dice}                                                        & \texttt{agri:Fisheries}                                                               & \begin{tabular}[c]{@{}l@{}}How many metric tons of river fish have been produced in Barguna district in \\ 2018-19?\end{tabular}                                                        & 5845  metric tons                                                                                           & 5845  metric tons                                                                                           & Y                                                                    \\ \hline
Q10                                                             &                                                                                & \texttt{agri:Production}                                                              & \begin{tabular}[c]{@{}l@{}}What is the name, scientific name, sowing time and harvest time of Garlic?\end{tabular}                                                             & Not Applicable                                                                                   & Can Give Answer                                                                                  & Y                                                                    \\ \cline{1-1} \cline{3-7} 
Q11                                                             &                                                                                & \texttt{agri:Fisheries}                                                               & What is the name and description of the fishery sector?                                                                                                                           & Not Applicable                                                                                   & Can Give Answer                                                                                  & Y                                                                    \\ \cline{1-1} \cline{3-7} 
Q12                                                             & \multirow{-3}{*}{Other Query}                                                         & \texttt{agri:Forestry}                                                                & What is the name and description of the forestry sector?                                                                                                                          & Not Applicable                                                                                   & Can Give Answer                                                                                  & Y                                                                    \\ \hline

Q13                                                             & \begin{tabular}[c]{@{}l@{}}Inter-cuboid\\ Query\end{tabular}              & \begin{tabular}[c]{@{}l@{}}\texttt{agri:Fisheries} \\ \texttt{agri:Production}\end{tabular} & \begin{tabular}[c]{@{}l@{}}District-wise amount of total production of crops and fisheries in different year\end{tabular}                                                      & Not Applicable                                                                                   & Can Give Answer                                                                                  & Y                                                                    \\ \hline

Q14                                                             &                                                                                & \begin{tabular}[c]{@{}l@{}}agri:Production\\ and Exiobase\end{tabular}       & \begin{tabular}[c]{@{}l@{}}Year wise Carbon-di-oxide emissions for different products\end{tabular}                                                                             & Not Applicable                                                                                   & See Section~\ref{sec:carbon}                                               & Y                                                                    \\ \cline{1-1} \cline{3-7} 
Q15                                                             & \multirow{-2}{*}{\begin{tabular}[c]{@{}l@{}}Federated\\ Query\end{tabular}}                                                                               & \begin{tabular}[c]{@{}l@{}}agri:Forestry\\ and Wikidata\end{tabular}       & \begin{tabular}[c]{@{}l@{}}District-wise comparison between current forest area and ideal ones.\end{tabular}                                                                             & Not Applicable                                                                                   & See Section~\ref{sec:forest}                                               & Y                                                                    \\ \cline{1-1} \cline{3-7} 
Q16                                                             &   & \begin{tabular}[c]{@{}l@{}}agri:Fisheries \\ and Wikidata\end{tabular}     & \begin{tabular}[c]{@{}l@{}}District wise comparative analysis between fish production and\\   population\end{tabular}                                                              & Not Applicable                                                                                   & See Section~\ref{sec:fish}                                                 & Y                                                                    \\ \hline
\end{tabular}%
}
\end{adjustbox}
\end{table*}

\subsubsection{Correctness}
BDAKG allows data-driven analysis through SPARQL queries, OLAP (intra and inter cuboid) queries, and federated data analysis (discussed in Section~\ref{section:analytics}).  Upon the execution of SPARQL/OLAP queries, we verify the accuracy of the results derived from it. To do so, we create a set of competency queries and categorize them based on different OLAP operations. The focused OLAP operations here are as follows: 1) Roll-up: This operation involves aggregating data from a specific level in a hierarchy to a higher level within the same hierarchy. Consequently, a roll-up results in reduced detail visibility. 2) Drill-down: It is the opposite operation of roll-up. Therefore, performing a drill-down operation reveals more detailed information. 3) Slice: It entails removing a dimension from a cube by fixing a single value at a particular level within that dimension. 4) Dice: Dice involves retaining the cells of a cube that satisfy a Boolean condition across dimension levels, attributes, and measures~\cite{vaisman2014data}. We also check whether BDAKG allows inter-cuboid queries and exploratory analytics. Equivalent SPARQL queries have been crafted to extract answers from the knowledge graph based on the specified competency questions. Users can access this collection of competency questions through a user interface accessible at \url{https://bdakg.netlify.app}. Users can pose these questions to the repository and receive answers. 
 
 Table~\ref{tab:competencyQuestions} shows queries, their type, answer received from original dataset, answer recieved from BDAKG, and the correctness of BDAKG. It is shown that both the original dataset and BDAKG correctly address queries Q1-Q9. However, BDAKG achieves results in less than one minute, whereas an expert user would typically spend over two hours for manual calculations. Unlike the original dataset, BDAKG can respond to queries Q10-Q12 due to its enrichment from various datasets, encompassing additional details such as harvest and sowing times, scientific names, and product descriptions. Furthermore, owing to its annotation with MD semantics, BDAKG can handle inter-cuboid queries (Q13). Its capacity to explore external knowledge graphs through links also enables it to respond to federated queries (Q14-Q16).

\subsection{Federated Data Analysis}\label{section:analytics}
The presence of ample agricultural data and the ability to explore through external links, BDAKG enables federated data analysis that can assist in making educated recommendations for decision makers. In the federated setting, a query initiator aims to respond to a data analysis query by engaging multiple data owners (datasets) who possess their respective local raw data. The actual raw data is not shared or transmitted; instead, intermediary query responses, intended for aggregation at the query initiator, are exchanged to fulfill the desired query~\cite{elkordy2023federated}. Our BDAKG  can be utilized to achieve 4 out of 17 of United Nations
sustainable development goals (SDG) \cite{fund2015sustainable}. They are zero hunger (SDG-2), decent work and economic growth (SDG-8), responsible consumption and production (SDG-12), and climate action (SDG-13). In the following subsections, we highlight some of the insights that we attained from our integrated knowledge graph using federated data analysis. These can utilized as recommendations for respective decision makers.

\subsubsection{Carbon footprint}\label{sec:carbon}
Carbon footprint is the amount of green house gases (such as Carbon dioxide) emitted in the making of an product or conduction of an activity. Carbon-di-oxide (CO$_2$) is one of the major green house gases that is contributing to global warming and eventually, climate change. BDAKG consists of Bangladesh's agricultural data which unfortunately does not contain Carbon footprint information of the products. However, Exiobase \cite{stadler2021exiobase} is a multi-regional database that provides carbon footprint details for various products. Its semantic version, available at \url{https://odas.aau.dk/}, is linked to BDAKG. The database is also accessible at \url{https://lca.aau.dk/FootprintAnalyser}. While Exiobase includes carbon footprint information by country, Bangladesh is not listed. In this case, we utilize the carbon footprint data of India for analysis since the product processing and harvesting mechanisms are similar.

\begin{figure}
    \centering
    
    \begin{subfigure}{0.46\textwidth}  
        \includegraphics[width=\linewidth]{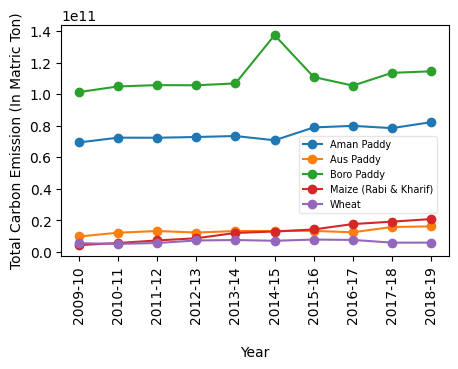}
        \caption{CO$_2$ emission for main cereals production.}
        \label{fig:subfig1}
    \end{subfigure}
    \hfill
    \begin{subfigure}{0.42\textwidth}
        \includegraphics[width=\linewidth]{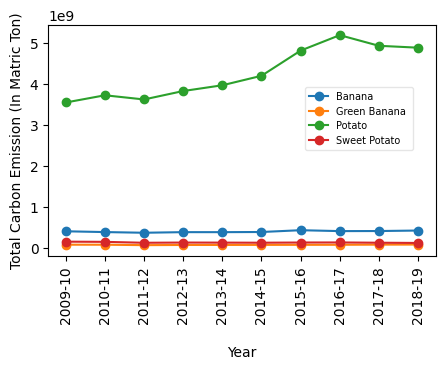}
        \caption{CO$_2$ emission for alternate carbohydrate production.}
        \label{fig:subfig2}
    \end{subfigure}
    \hfill
    
    \begin{subfigure}{0.45\textwidth}  
        \includegraphics[width=\linewidth]{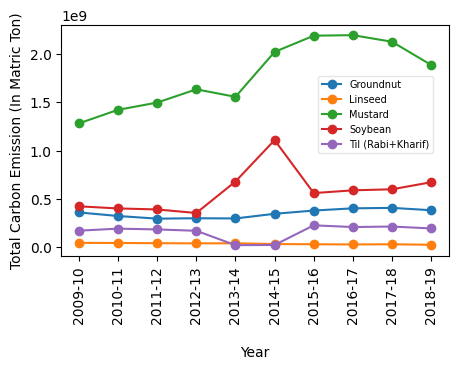}
        \caption{CO$_2$ emission for oilseeds production.}
        \label{fig:subfig3}
    \end{subfigure}
    \hfill
    \begin{subfigure}{0.47\textwidth}
        \includegraphics[width=\linewidth]{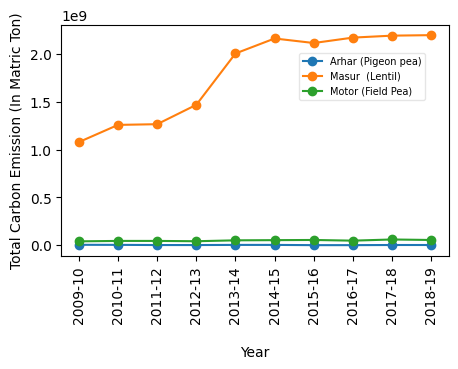}
        \caption{CO$_2$ emission for pulses and lentils production.}
        \label{fig:subfig4}
    \end{subfigure}

    \label{fig:overall}
 \hfill
   \begin{subfigure}{0.44\textwidth}
        \includegraphics[width=\linewidth]{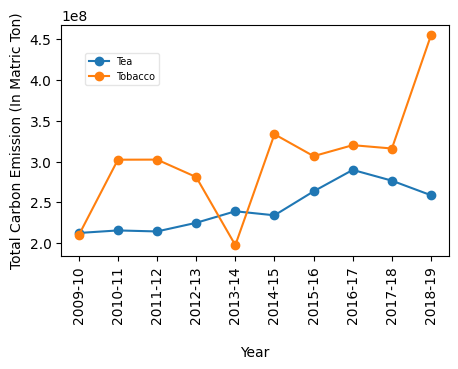}
        \caption{CO$_2$ emission for bevarage production.}
        \label{fig:subfig5}
    \end{subfigure}
    \hfill
       \begin{subfigure}{0.44\textwidth}
        \includegraphics[width=\linewidth]{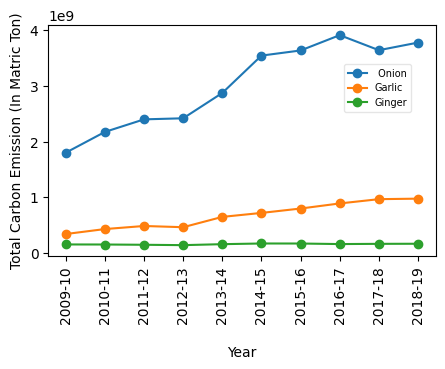}
        \caption{CO$_2$ emission for spices' production.}
        \label{fig:subfig6}
    \end{subfigure}

    \hfill
     \caption{Comparative analysis of the total CO$_2$ emissions across various categories from the fiscal year 2009-10 to 2018-19.}
\label{fig:carbon}
\end{figure}

\begin{figure}[h!]
    \centering
    \includegraphics[scale = 0.38]{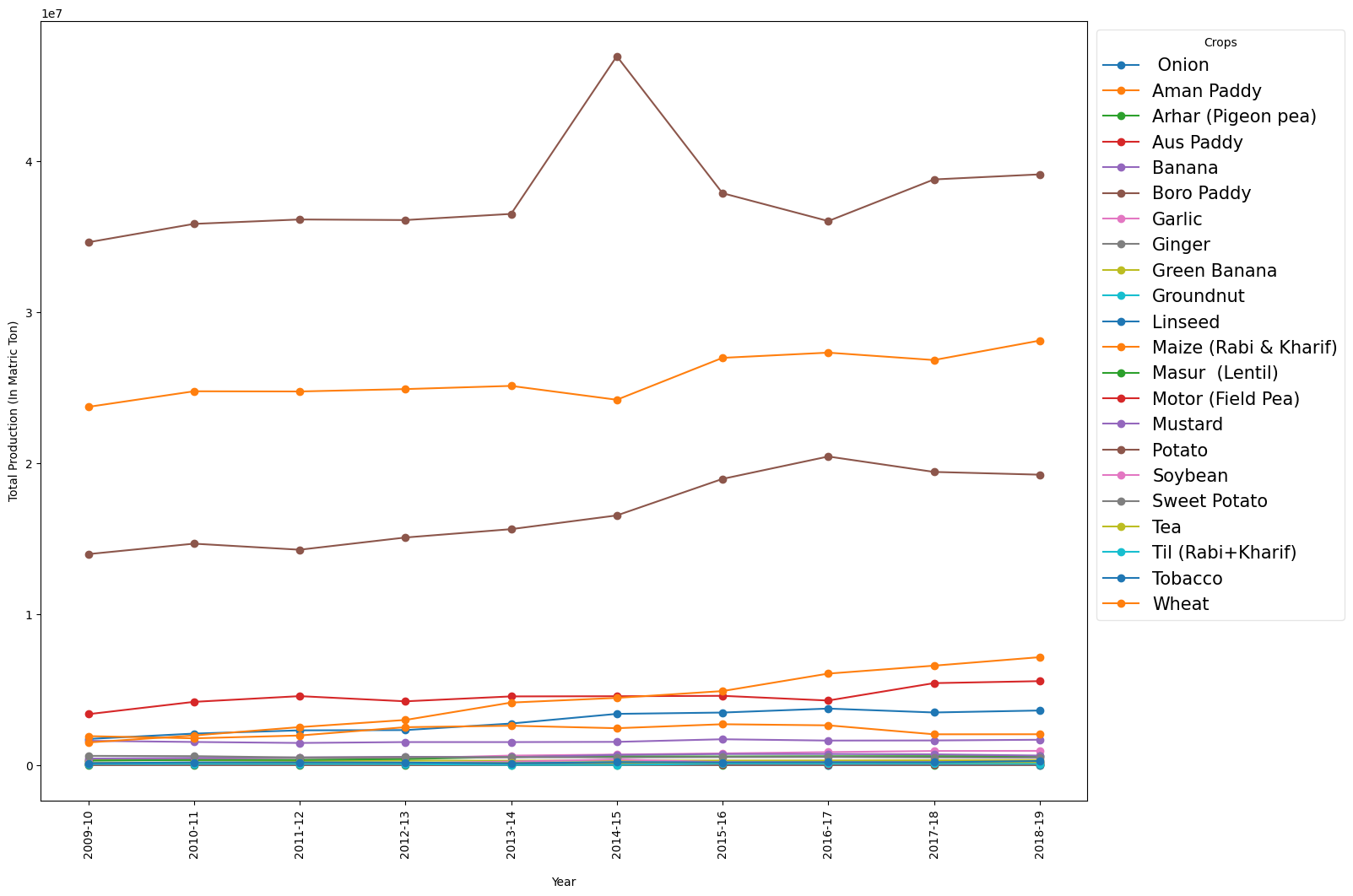}
    \caption{Production trend of various crops in ten fiscal years}
    \label{fig:crop_production_linechart}
\end{figure}

\begin{figure}[h!]
    \centering
    \includegraphics[scale = 0.55]{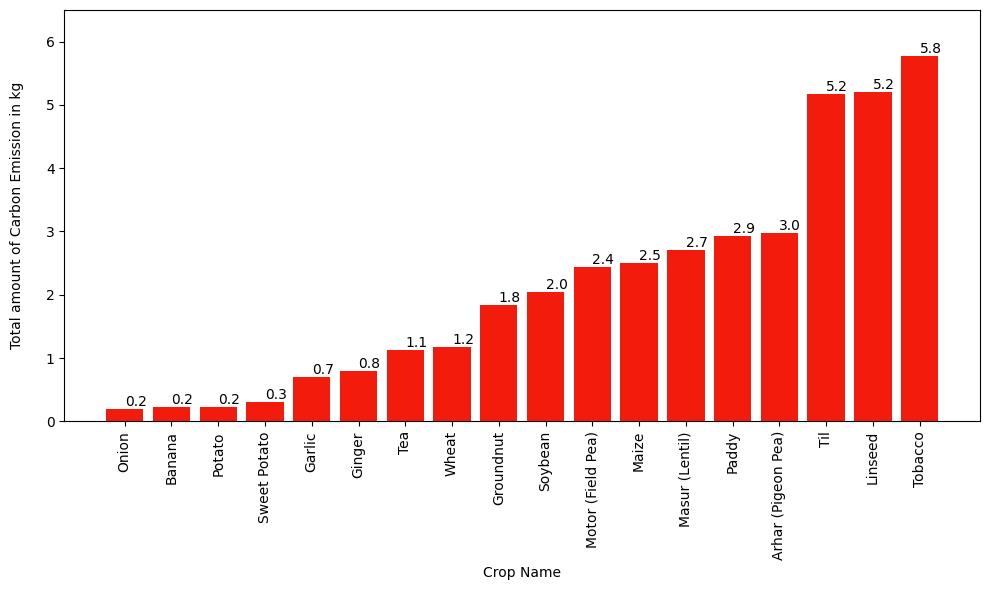}
    \caption{CO$_2$ emissions (in kg) per kg of different crop production (Kg CO$_2$/ Kg)}
    \label{fig:carbonemmission}
\end{figure}

Figure~\ref{fig:carbon} depicts a comparative analysis of the total CO$_2$ emissions across various categories from the fiscal year 2009-10 to 2018-19. Figure~\ref{fig:subfig1} illustrates CO$_2$ emision for main cereals production. It can be seen that overall paddy has a high (CO$_2$) emission rate. That has coupled with the high production of paddy and contributed to boosting the total emission too. The high production is due to the fact, that paddy is the source of rice, the main staple food of Bangladesh. Rice in Bangladesh, is also the major source of carbohydrate too. Our analysis found Aus paddy to have lowest production among the three kinds of paddy. Yet, it yields higher than other significant crops, as can be seen in Figure~\ref{fig:crop_production_linechart}. The trend is a rising one. Among other cereal crops produced in Bangladesh, wheat has low emission rate (see Figure~\ref{fig:carbonemmission}). Therefore, it can be recommended that the government of Bangladesh should raise awareness to gradually reduce rice consumption and paddy production. As replacement, wheat production can be increased. Wheat based bread is already a popular cuisine in many parts of Bangladesh, especially as breakfast item. A good proportion of health conscious people - elderly and young alike - are already building the habit of having wheat based bread for lunch and dinner. Therefore, wheat can be a viable replacement for paddy. But a matter of concern is the eminent declining trend of wheat production, as can be viewed from Figure \ref{fig:subfig1}. This should be countered by taking measures to increase wheat production. \par

However, it should be mentioned that alternate sources of carbohydrate can be found among fruits and fodder crops like potato, sweet potato, and banana, as seen from Figure~\ref{fig:subfig2}. Potato is the staple food in countries such as Ireland, Peru, Bolivia, and Russia. Sweet potato is staple source of carbs in Papua New Guinea, Uganda, and parts of China and Japan. Similarly banana is a major staple food in west and central Africa, the Caribbean islands, Central America and Northern South America. Among the three while banana has lowest CO$_2$ emission, sweet potato is lowest in total emission. Irrespective of this, it is notable that these crops have low emission rates and total emissions, when compared to the cereal crops. Roasted sweet potato is already popular as a snack in Bangladesh, whereas potato is popular both in its mashed form (known as '\textit{Aloo Bharta}') and as a curry vegetable. It can be seen from Figure \ref{fig:subfig2} that banana yield has been in a slightly increasing trend. This should be maintained and enhanced. However, sweet potato production did not cross even a million metric ton in last ten years. To utilize sweet potato as a viable substitute food source that emits less CO$_2$, its production should be increased. 

From Figure~\ref{fig:subfig3}, it can be seen that among oilseeds, soybean has considerably very less CO$_2$ emission. Soybean oil is largely consumed in Bangladesh. However, Bangladesh heavily depends on importing for meeting the demand for soybean oil. Since soybean has such less CO$_2$ emission, its production can be increased so that soybean oil can be manufactured within the country, rather than being imported. Figure~\ref{fig:subfig4} shows that among pulses and lentils, masur lentil has highest emission amount. Lentils are a significant source of protein in Bangladesh, especially masur. To reduce contribution to global warming, arhar can be a viable replacement. \par
Among the beverage and narcotic herbs, it can be observed from Figure~\ref{fig:subfig5} that tobacco exhibits a relatively higher CO$_2$ emission rate and total emission compared to tea. This observation should be taken into consideration alongside the hazardous impact of tobacco products such as cigarettes on public health. The production of Tobacco should hence be reduced and that of tea can be increased to substitute it. 
Figure~\ref{fig:subfig6} illustrates that onions release a higher amount of CO$_2$ compared to garlic and ginger. Specifically, in the fiscal year 2018-19, onions emitted 2.5 and 3.5 times more CO$_2$ than garlic and ginger, respectively. Meanwhile, in Bangladesh, the cost of onions is steadily rising, prompting a suggestion to consider reducing the daily consumption of onions.

\par
We have just mentioned the potential of wheat, soy beans, arhar, tea, and ginger for an eco-friendly agriculture. However, Figure~ \ref{fig:crop_production_linechart} shows that they have had very less production in last few years. To capitalize their benefits, their production should be boosted.

\subsubsection{Forest area}\label{sec:forest}
To meet ecological balance, at least 25\% of a country should be forest area. With BDAKG, we broke it down at district level to observe districts that have insufficient forestry, as per the above mentioned metric. Since BDAKG is a knowledge graph, we were able to use the power of federated query in this analysis. Using this technique, BDAKG can incorporate additional information about its attributes that are not directly available in it. It does so, by being linked to prominent knowledge graphs such as Wikidata and Dbpedia, among others. BDAKG does not contain information on district area, For forest area analysis, it brought the data from Wikidata through a federated query. Table \ref{tab:forest_insight} depicts the findings.\par

\begin{table*}[]
    \centering
    \begin{tabular}{|c|c|c|}
    \hline 
    \textbf{District} & \textbf{Forest area it has} &\textbf{Forest area it should have}\\
    \hline
    Barguna&75000&113131.5\\
    \hline
    Habiganj&36360.7&162879\\
    \hline
    Kurigram&128.59&138690.25\\
    \hline
    Nilphamari&1200.08&95543\\
    \hline
    Patuakhali&150000&199000.75\\
    \hline
    Pirojpur&6000&78938\\
    \hline
    Sunamganj&18012.3&232811.5\\
    \hline
    \end{tabular}
    \caption{Districts that have insufficient amount of forest}
    \label{tab:forest_insight}
\end{table*}

The table lists districts that do not have sufficient forest area. It can be observed that among them are included northern districts such as Kurigram and Nilphamari. Northern Bangladesh is prone to drought. Thus government should focus on increasing the forestry in these two districts, alongside others to tackle drought and other climate calamities.

\begin{figure*}[h!]
\centering
\includegraphics[scale=.34]{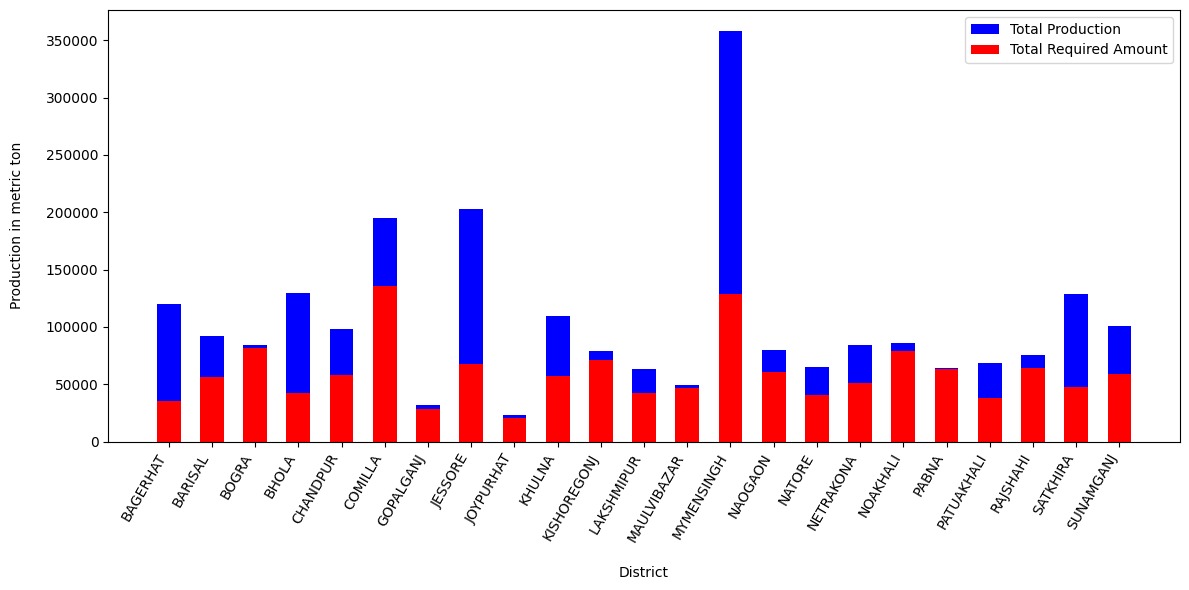}
    \caption{ Districts with Fish Production Exceeding Demand. }
    \label{fig:fish}
\end{figure*}

\subsubsection{Fish sector}\label{sec:fish}
Bangladesh has emerged as one of the world's leading fish producers after becoming self-sufficient in fish production. Bangladesh's export revenues are mostly dependent on its fisheries resources, which account for the largest portion of export revenue \cite{shamsuzzaman2020economic} compared to agriculture. Over the past 15 years, aquaculture has grown in importance as one of Bangladesh's three most diverse fisheries resource sectors, along with inland open, marine, and cultural. By 2026 \cite{islam2018trends}, it is expected that Bangladesh would move up from the LDC to become a mid-income country. In that scenario, the export of fish and related goods can play a key role, like the pharmaceutical and garments sectors. Bangladesh is currently one of the few LDCs that has been granted permission to export fish goods to the EU \cite{golub2014fishing}. Bangladesh earned approximately US\$533 million in the fisheries sector in the fiscal year 2021–2022, accounting for more than 1\% of the nation's total export revenue and 3.57\% of its GDP \cite{Introduc71:online}. It enriches our economy by increasing foreign exchange and employing more than 19 million people, directly or indirectly. The Bangladesh government has set a target of 8.5 million tonnes \cite{Introduc71:online} of fish production by 2041 to make Bangladesh a Smart Bangladesh as well as enriching the Smart Economy. To achieve this goal, the industry must experience a remarkable surge in fish production, which can only be referred to as "\textit{Smart Fish Farming}" \cite{yang2021deep}—a word that combines modern farming methods with technology. Encouraging sustainable resource management is one of the main goals of smart fishing practices. Fishermen can find regions with abundant fish populations and steer clear of over fished areas by using data-driven decision-making. \par

The recommended dietary intake for fish is 60 gram per capita per day, for Bangladeshi population \cite{faruque2013desirable}. On a yearly basis, this is 21.9 kg per capita per year. We used this figure and found some interesting patterns in BDAKG. We used the power of federated query to bring population data of each district from Wikidata and multiplied  0.0219 with the population. This gave us an estimate about required production (in metric ton) threshold of that district. We found that in 28 out of 64 districts in 2019-20 (i.e., latest fiscal year available), just inland open and closed fish production exceeds the required amount, as can be seen in Figure~ \ref{fig:fish}. The excess fish produced in these districts is surplus and can therefore be exported abroad. 



\section{Related work}\label{sec:rw}
Agriculture is the main and primary driving force of any civil institution, be it family, society, or state. It becomes even more impactful in a river-laced country like Bangladesh, comprising 42.7\% of the country's workforce \cite{Banglade17:online}. Therefore, ample agriculture data is generated both from government and non-government sources at regular periods. To present the information hidden in such data in a useful way to decision makers, Business Intelligence (BI) tools are necessary. Such information creates a picture of the past and present scenario of the agricultural landscape. On the other hand, to make real-time future predictions and reliable prescriptive recommendations from this information, analytical approaches such as data mining prove very fruitful. In light of this understanding, previous works can be categorised into two broad categories. 

\begin{itemize}
    \item [-] First category contains those works that explore the application of BI tools such as data warehousing and OLAP operations to agriculture data.
    \item [-] Second category of works applied data mining techniques to extract insights and patterns from agricultural datasets.
\end{itemize} 
They are briefly discussed below. \par

\subsection{Research focusing on application of BI tools and techniques to agriculture data}
Studies listed under this category investigate application of BI to draw out meaningful and actionable insights from agriculture data. In this pursuit, state-of-the-art knowledge representation techniques and data analysis tools, including data warehouse and knowledge graphs are explored in them. \par
In \cite{tyrychtr2015evaluation}, authors evaluated the state of BI application to small farms in the Czech Republic. They used data of 135 farms from various regions of the country. The total land size of the farms amounted to 500 hectares. They have shown that despite inventment in information technology among farmers, BI tools and techniques have not been implemented much in their agriculture. Although they highlighted this discrepancy regarding BI application in agriculture in the Czech farms, they did not model or implement any BI tool such as data warehouse or data mart in their paper.

Maliappis et al. \cite{maliappis2015online} provided a layout of a DW for Greek agricultural data and minimal OLAP implementation. They collected their data from Hellenistic Statistical Authority (ELSTAT) and Farm Accountancy Data Network (FADN/RICA). However, they have limitations in that the ELSTAT data has early historical data unavailable. Also, the administrative division information of Greece underwent multiple changes leading to inter-temporal differences.

Radulescu e al. \cite{radulescu2009multidimensional} presented a multidimensional data cube called CUBECTH for agriculture. It offers farmers solutions and responds to their ad hoc queries utilising the star schema. The model facilitates OLAP operability. Authors intended the study to be production management oriented, rather than business oriented. However, they are limited by the fact that they did not make their dataset available online.

In \cite{wisnubhadra2020open}, a new spatiotemporal data warehouse was proposed that integrated service-oriented architecture and open data sources. They used data from Village and Rural Area Information System (SIDeKa). It contains daily agricultural production transaction data. Although the open data they used is available as linked open data, constructs of the new data warehouse they formed have not been made available as linked open data.\par
The works mentioned above have the recurring limitation that their data is not made available as semantic linked data.

\subsection{Research focusing on applying data mining techniques on agriculture data:}
Agriculture data contain abundant information that can be transformed into valuable insights. This can be done using data mining techniques. These techniques extract recurring patterns in the data  that can later be utilised to discover new knowledge.

In \cite{Harsanyi2023DataMA}, they created a method for managing agriculture sustainably in the face of climate change using Data Mining and Machine Learning algorithms; however, they only assessed maize yield forecasting in a sample of 98 observations from a historical time series spanning 1921 to 2018 rather than relating other crop data like potato, amon, onion, etc..

Shakoor et al. \cite{shakoor2017agricultural} used  supervised machine learning techniques to build an intelligent information prediction analysis of farms in Bangladesh. They took into account six major crops over the past twelve years. 
In \cite{rahman2014application, majumdar2017analysis}, a comprehensive analysis was conducted using data mining tools to predict Bangladesh's rice yield, namely Boro, Aman and Aus, along with their subvarieties to analyse agricultural data and discover the best parameters to maximise crop yield. On the other hand, authors of \cite{sarker2012exploring,basak2010assessment} present the relationship between rice yields of three major rice crops (e.g., Aus, Aman, and Boro) and climate variables using aggregate-level (year-wise Bangladesh agriculture data) time series data for the 1972–2009 period using regression models and Decision Support System for Agro-technology Transfer (DSSA) model. Though in each of these studies, authors made remarkable contribution, they are not devoid of limitations. One such limitation is that they only focused on a few major crops at coarse, aggregated levels of Bangladesh agriculture data. They did not work with up-to-date minor crops or disaggregated (District Level) data in Bangladesh. The data they used was not published using FAIR and  linked open data (LOD) principles. They did not enable semantic linking of data sources and were not connected to agricultural external data sources. So it is impossible to infer more knowledge about agriculture to find the maximum outcome.

The current study is inspired by the new initiatives such as Open Government and Open Data following the 5* and FAIR data principles. 
Following is a summary of our contributions to address the limitations mentioned above. We create BDAKG, a multidimensionally annotated knowledge graph. It enables descriptive analysis using the multi-dimensional OLAP analysis technique to find the  crops (major and minor), fisheries, and forestry data at the granular level district-wise. New insights can be inferred from BDAKG using SPARQL queries. Moreover, it is built in a fashion that can be inorporated into the LOD cloud. To build BDAKG, we use the constellation schema to represent and analyse our agriculture data in a more flexible and normalised way compared to the star and snowflake schemas. We use the most up-to-date data available in BBS  \cite{Bangladesh:online}. We also enrich the concept of agriculture intelligence by integrating multiple sources of agricultural data, which is the combination of business intelligence and other types of information related to agriculture.

\section{Conclusion and Future work}\label{sec:con}
In response to the ongoing open data trend, organizations like the Bangladesh Bureau of Statistics routinely collect and publish agricultural data on the Web. However, the current dataset lacks suitability for interactive analysis and integration with other datasets. Addressing these challenges, we present the creation of a knowledge graph named BDAKG, annotated semantically and multidimensionally using the RDF, RDFS, OWL, and QB4OLAP vocabularies. The step-by-step process of generating this 44 MB knowledge graph, comprising approximately 382,000 RDF triples and linked to 363 external resources, is detailed. BDAKG is assessed for ETL performance, quality, and exploratory analytics, with the knowledge graph generation taking around 8.5 hours. The evaluation reveals that BDAKG is complete, correct, adheres to OLAP and FAIR principles, and is analytically exploitable. Insights related to reducing CO$_2$ emissions, promoting economic growth, and fostering sustainable forestry are drawn  through analysis and exploration of BDAKG via links.

Looking ahead, we plan to employ data mining techniques for forecasting and pattern extraction. Our goal is to extend our efforts to manage and publish data from various sectors in Bangladesh, including socio-economic, healthcare, education, energy, and the environment. Exploring these datasets aims to contribute to the development of a sustainable data infrastructure for open data in Bangladesh.



\bibliography{mybibfile}

\end{document}